\documentclass[aps,twocolumn,superscriptaddress,showpacs]{revtex4}

\usepackage{epsfig}
\usepackage{amsfonts}
\usepackage{amssymb}
\usepackage{mathrsfs}
\usepackage{theorem}
\usepackage{amsmath}
\usepackage{times}
\usepackage{color}

\usepackage{ifpdf}
\ifpdf
\usepackage{epstopdf}   
\fi

\newtheorem{theorem}{Theorem}
\newtheorem{defin}{Definition}
\newtheorem{lem}{Lemma}

\newcommand{\ket}[1]{\ensuremath{\vert#1\rangle}}
\newcommand{\bra}[1]{\ensuremath{\langle #1\vert}}

\newcommand{\kb}[2]{\ensuremath{\vert #1 \rangle \langle #2 \vert}}

\renewcommand{\vec}[1]{\ensuremath{\mathbf{#1}}}

\newcommand{\tr}{\ensuremath{\mathrm{tr}}}
\newcommand{\Span}{\ensuremath{\mathrm{span}}}
\newcommand{\Dim}{\ensuremath{\mathrm{Dim}}}
\newcommand{\wt}{\ensuremath{\mathrm{wt}}}

\def\id{\mbox{\small 1} \!\! \mbox{1}}

\def\id{\mbox{\small 1} \!\! \mbox{1}}

\def\id{{\mathchoice {\rm 1\mskip-4mu l} {\rm 1\mskip-4mu l} {\rm 1\mskip-4.5mu l} {\rm 1\mskip-5mu l}}}

\newcommand{\hus}{\textcolor{black}}

\begin{document}
\title{Magic state distillation in all prime dimensions using quantum Reed-Muller codes}

\author{Earl T. Campbell}
\email{earltcampbell@gmail.com}
\affiliation{Dahlem Center for Complex Quantum Systems, Freie Universit{\"a}t Berlin, 14195 Berlin, Germany.}
\author{Hussain Anwar}
\affiliation{Department of Physics and Astronomy, University College London, Gower Street, London WC1E 6BT, United Kingdom.}
\author{Dan E. Browne}
\affiliation{Department of Physics and Astronomy, University College London, Gower Street, London WC1E 6BT, United Kingdom.}

\pacs{03.67.Pp,03.67.Lx}

\begin{abstract}


We propose families of protocols for magic state distillation -- important components of fault tolerance schemes --- for systems of odd prime dimension. Our protocols utilize quantum Reed-Muller codes with transversal non-Clifford gates.  We find that, in higher dimensions, small and effective codes can be used that have no direct analogue in qubit (two-dimensional) systems.  We present several concrete protocols, including schemes for three-dimensional (qutrit) and five-dimensional (ququint) systems. The five-dimensional protocol is, by many measures, the best magic state distillation scheme yet discovered.  It excels both in terms of error threshold with respect to depolarising noise ($36.3\%$) and the efficiency measure know as ``yield'', where, for a large region of parameters, it outperforms its qubit counterpart by many orders of magnitude. 

\end{abstract}

\maketitle  

The central challenge of implementing scalable quantum computing is to protect quantum systems against noise and decoherence while retaining the capacity to perform computation.  Quantum error correction and fault tolerant techniques provide a solution to this problem, and a variety of constructions for fault tolerant quantum computation have been proposed~\cite{shor1996fault,Kit03,Rauss07,Knill04}. In all these schemes, a delicate balance must be maintained between coherently manipulating the encoded system while preserving the protected subspace and prohibiting the proliferation of errors.  For example, for schemes built on stabilizer codes \cite{G02a} transversal gates have the desired properties, while in topological systems, topologically protected braiding operations \cite{Kit03} provide the logical gates. While much work in quantum computation has focussed upon qubits (two-level systems), it is known that for any prime $d$, effective codes exist for storing $d$-level quantum systems~\cite{G02a,G01a,Aharonov}.  Thus qudit systems are also candidates for scalable fault tolerant quantum computation. 

In many approaches, the protected unitary gates are  a subset of the so-called Clifford group. The stabilizer operations (comprising Clifford unitaries as well as preparation and measurements in the computational basis)  are known to be efficiently classically simulatable~\cite{G01a,G02a,denNest}, and  on their own are not universal for quantum computation. Furthermore, several theorems have shown~\cite{Xie08,Zeng11,Eastin09, Bravyi12} that, in general, there is a tension between providing protection against generic noise and achieving universal quantum computing.  

Despite these obstacles, fault tolerant universal quantum computing is possible \cite{shor1996fault}.  One particularly successful approach, known as state-injection, is to achieve universality by augmenting the fault tolerant operations with a supply of many copies of a suitable ancillary resource state.  While methods for direct preparation of sufficiently noise-free  protected resource states have been proposed  \cite{shor1996fault}, a particularly elegant solution can be provided by distillation techniques, where many noisy copies of a resource state can be distilled to arbitrary fidelity by using only error-protected operations, while preserving the error threshold of the model. 

\begin{figure}
\includegraphics{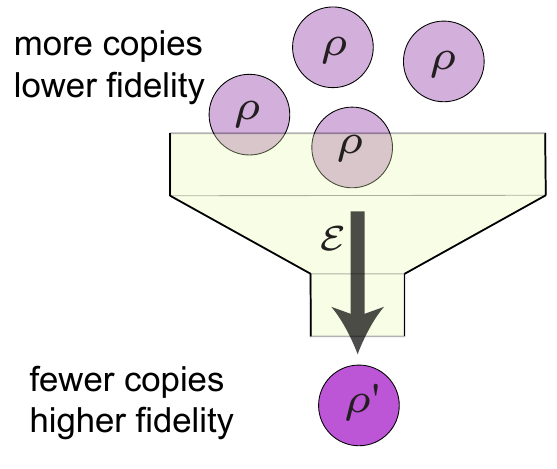}
\caption{An outline of a round magic state distillation protocol.  Within many architectures of fault tolerance quantum computing a large proportional of the device is committed to these magic state factories.  Each attempt uses $n$ copies of a state $\rho$, and when successful outputs a state $\rho' \propto \mathcal{E}(\rho^{\otimes n})$. Given $n$ successful attempts, these are used as inputs into the next iterate.  Within the magic states model the completely positive map, $\mathcal{E}$, is composed of a sequence of Clifford unitaries and Pauli measurements.  This figure illustrates a protocol where $n=4$, for example the ququint, $d=5$, protocol that we discuss throughout the article.}
\label{Fig_iterative}
\end{figure}

Here we consider the typical but idealized case, where the available protected operations are perfect stabilizer operations.  As such whenever we speak of a qubit we mean a qubit encoded into either a topological system or stabilizer code that provides protection of stabilizer operations.  This has become known as the magic states model and was first studied by Bravyi and Kitaev~\cite{BraKit05} who proposed two protocols for distillation of qubit magic states, the resource for state-injection of a non-Clifford unitary.   In parallel, Knill proposed the concept of a post-selected quantum computer~\cite{Knill04b,Knill04,Knill05} that used state preparation protocols that appeared distinct to magic state distillation,  but were later shown to be equilvalent~\cite{Rei01a}.  These techniques are a key component of many fault tolerance schemes, including for example the topological cluster state scheme~\cite{RHG01a,Rauss07,Devitt09,Fowler09,Barrett10,Benj10,Benj12}.

Additional protocols were later discovered by Reichardt~\cite{Rei01a,Rei02a,Rei03a}, which increased the family of qubit states known to be distillable.  Conversely, Campbell and Browne~\cite{Camp10a,Camp09c} showed that no finite iterative protocol could distill all mixed nonstabilizer states.  Many other results have contributed to our understanding of the magic states theory for qubit systems~\cite{VirPlen,Howard09,Virmani10,Anderson12} and a 5-qubit distillation protocol has been implemented in an NMR system~\cite{MagicNMR}.

The theory of higher dimensional quantum computation~\cite{Bartlett02,Zhou03,Bremner1,Bremner2}, stabilizer operations and error correcting codes~\cite{Gott99,Beaudrap} is well known.  However, higher dimensional magic state models have been largely neglected until recently.  In anyonic systems the dimensionality of the available stabilizer operations is determined by the underlying physics, and so with some physical systems we would have no choice but to work in the higher dimensional model (see e.g.~\cite{Wootton,Shtengel12}).  Recent progress on this problem has centered on exploiting a discrete phase space, or Wigner function, representation of quantum states~\cite{Gross06,Howard11,Veitch}.  Notably, Veitch \textit{et al}.~\cite{Veitch} showed that states with positive Wigner functions can never be used as a resource for magic state distillation.  Although all stabilizer states have positive Wigner functions, there also exist undistillable nonstabilizer states and so bound magic states.   

These developments on no-go results took place without any known distillation protocols in higher dimensions.  However, recently we have proposed a protocol for 3-dimensional, qutrit, systems that uses a generalization of the 5-qubit code~\cite{Hussain1}.  Magic state distillation was observed there, but the error suppression was slower than in qubit protocols.   Here we present a family of protocols that distil magic states in any odd prime dimension and do so with a quadratic reduction in noise per iteration.  As such the protocols are competitive with (and in some cases outperform) the best previously-known qubit protocols. 

Our protocols exploit higher dimensional quantum Reed-Muller codes~\cite{Pradeep2005} and so generalize the qubit protocol of Bravyi and Kitaev~\cite{BraKit05} that used a 15-qubit quantum Reed-Muller code.  This 15-qubit code was, to our knowledge, first developed by Knill, Laflamme and Zurek~\cite{Knill96} and later developed by Steane~\cite{Steane99}.  These quantum codes are constructed from classical Reed-Muller codes~\cite{Sloane,AssmusKey,RM68,RM68b,RM70,RM98}, which have played a pivotal role in classical coding theory.  Notably, the family of Reed-Muller codes includes the infamous Reed-Solomon code used for communication with the Voyager space probe and data storage on compact disks.

We begin with a formal description of the Clifford group and the magic states model.  This allows us to state our main theorem; roughly that magic state distillation is possible in higher dimensions.  Next we review some basic theory of quantum error correction and show what properties of error correcting codes would enable us to build a protocol for magic state distillation.  This sets the stage for constructing codes, the quantum Reed-Muller codes, which have the required properties.  Next we introduce some additional tools from classical coding theory that helps to simplify our analysis for a uniform depolarizing noise model.  

We consider several measures of performance for many protocols applied to systems of up-to 19 dimensional systems.  The measures indicate that two protocols for qutrits  (3-dimensional systems) and ququint (5 dimensional systems) perform well compared to both qubit protocols and protocol for even higher dimensional systems.  For these two protocols we investigate their performance in more detail.  We find that the qutrit protocol performs well, but that the ququint outperforms all other known magic state protocols, both, in terms of the degree of error on the initial state it can tolerate and the efficiency of the protocol.   We will see that the effectiveness of these protocols can be related to properties of the Clifford group.   The startlingly good performance of these protocols make higher dimensional systems a enticing alternative to qubit systems.

Finally, we show how to perform state-injection to convert the distilled magic states into non-Clifford gates. The addition of any non-Clifford unitary to the set of n-qudit Clifford gates gives a set of gates dense in $\textrm{SU}(d^{n})$ and so approximately universal via the Solovay-Kitaev theorem. This fact is well-known for qubits, but is also true for general prime $d$, which follows from theorems proven by Nebe, Sloane and Rains in their recent book~\cite{nrsbook} (see Appendix~\ref{universality}).  

\section{Stabilizer operations and the magic states model}

We are interested in $d$-dimensional quantum systems, or qudits, where $d$ is an odd prime.  The computational basis states are labeled by $j \in \mathbb{F}_{d}$, where $ \mathbb{F}_{d}$ denotes the finite field of $d$ elements.  For such systems, the so-called Pauli group, $\mathcal{P}_{d}$, is generated by
\begin{eqnarray}
	X & = & \sum_{j \in \mathbb{F}_{d}}  \kb{j \oplus 1}{j} , \\ \nonumber
	Z & = & \sum_{j \in \mathbb{F}_{d}} \omega^{j} \kb{j}{j} ,
\end{eqnarray} 
where $\oplus$ is addition modulo $d$, and $\omega = \exp ( i 2 \pi / d )$.  The conjugation relation, $X Z= \omega^{-1} Z X$, is easy to verify and used throughout.   The Pauli group over $n$ qudits,  $\mathcal{P}_{d}^{n}$, is the $n$-fold tensor product of the single qudit Pauli group.  Consider an Abelian subgroup of the Pauli group, $\mathcal{S}$, which contains the identity but no other multiple of the identity, e.g. $\omega \id \notin \mathcal{S}$.  Associated with this group is a physical subspace, called a stabilizer code, and a projector onto this subspace, $\Pi \propto \sum_{s \in \mathcal{S}} s $.  We equate $\Pi$ with the code and call the group $\mathcal{S}$ the stabilizer of the code.  When the code is $1$-dimensional, the projector describes a pure quantum state, which we call a pure stabilizer state.  We will also follow common terminology and call any probabilistic ensemble of pure stabilizer states a stabilizer state, even when there does not exist a unique stabilizer group describing the mixture.  


The Clifford unitaries $\mathcal{C}_{d}^{n}$ are those that conjugate the Pauli group to itself, so $\mathcal{C}_{d}^{n} = \{ U ; U \mathcal{P}_{d}^{n} U^{\dagger}=\mathcal{P}_{d}^{n} , U \in U(d^{m}) \}$.  The whole Pauli group is a subgroup of the Clifford group, $\mathcal{P}_{d}^{n} \subset \mathcal{C}_{d}^{n}$.  Gottesman~\cite{Gott99} introduced several other Clifford gates, including the single qudit gates
\begin{equation}
\label{eqn_Cliffords}
\begin{array}{rcl}
P & = & \sum_{j}  \omega^{  j(j-1) / 2 }  \kb{j}{j} , \\
H & = & ( \sum_{j,k}  \omega^{  jk } \kb{j}{k} )/ \sqrt{d} ,
\end{array}
\end{equation}
and the 2-qudit gate, $ SUM$ (or generalized CNOT) gate,
\begin{equation}
	SUM=\sum_{j}  \kb{j}{j} \otimes X^{j} .
\end{equation}
These gates have been shown to generate the whole Clifford group \hus{\cite{Clark06}}.  The magic states model also allows the implementation of so-called Pauli measurements.  Given any Pauli $U \in \mathcal{P}_{d}^{n}$, which we express as $U = \sum_{k=0}^{d-1} \omega^{k} U_{k}$,  we allow for POVM measurements with elements $\{ U_{k} \}$.  It is commonplace, though a modest abuse of terminology, to speak of measuring the Pauli $U$.   

For an $n$-qudit system the space of possible density matrices is within the set of bounded operators,  $B(\mathcal{H}^{d^{n}})$, acting on $\mathcal{H}^{d^{n}}$.  For such a space the set of physical stabilizer operations allowed in the magic states model is captured by the following.
\begin{defin}
Consider a completely positive map $\mathcal{E} :   B(\mathcal{H}^{d^{n_{\mathrm{in}}}}) \rightarrow  B(\mathcal{H}^{d^{n_{\mathrm{out}}}})$.  The map is a stabilizer operation if and only if it can be composed from the following:
\begin{enumerate}
	\item Clifford unitaries;
	\item measurements and subsequent projections on stabilizer subspaces;
	\item preparation of fresh ancilla in a stabilizer state;
	\item tracing out of unwanted qudits;
	\item adaptive decision making based on both measurement outcomes and random coin tosses.
\end{enumerate}
\end{defin}
The number of qudits output and input may differ, as is typically the case when magic state distillation is performed.

\section{M-type gates and M-distillation codes}

Every code defines an iterative scheme for magic state distillation. However, some codes are much more suitable than others, and their usefulness can often be inferred from abstract properties of the code. In particular, the 15-qubit Reed-Muller code exploited by Brayvi and Kitaev has a very special property. There exists a product operator, of the form $U^{\otimes n}$, that acts on the logical basis as a non-Clifford operator.  Such a code is said to have \textit{transversal} non-Clifford gates, and we will consider generalizations of the qubit Reed-Muller codes with this remarkable property.   

The transversal non-Clifford gate of the 15 qubit code, the so-called $\pi/8$ gate denoted $U_{\pi/8}$,  has another additional interesting property; For all Pauli $P \in \mathcal{P}_{2}^{n}$, we have that $U_{\pi/8} P U^{\dagger}_{\pi/8} \in \mathcal{C}_{2}^{n}$.  Gottesman and Chuang defined this set of gates as the second level of an infinite hierarchy of qubit gates~\cite{CliffHier}.  The hierarchy generalizes easily to quqits.
\begin{defin}
\label{Def_Hier}
The $k^{\mathrm{th}}$ level of the Clifford hierarchy for $n$ quqits is the set
\begin{equation}
	\mathcal{C}_{d}^{n}(k) = \{  U |  \forall P \in \mathcal{P}_{d}^{n} , U P U^{\dagger} \in \mathcal{C}_{d}^{n}(k-1)   \} ,
\end{equation}
where the bottom level is the Pauli group $\mathcal{C}_{d}^{n}(0)=\mathcal{P}_{d}^{n}$.
\end{defin}
The hierarchy is defined recursively with the $k^{\mathrm{th}}$ level as the set of unitaries that conjugate the Pauli operators to a unitary in  the ($k-1)^{\mathrm{th}}$ level.  The bottom level is fixed as the Pauli group and the first level is simply the Clifford group $\mathcal{C}_{d}^{n}(1)=\mathcal{C}_{d}^{n}$.  Whereas higher levels are not groups.   The quqit gates of interest share these properties with the qubit $U_{\pi/8}$ gate and are defined as follows.
\begin{defin}
\label{Def_Mgate}
The set of gates $\mathcal{M}_{d}^{m}$ contains all $M$ such that:
\begin{enumerate}
	\item $M$ is diagonal in the computational basis;
	\item $M^{d^{m}}= \id$;
	\item $M \in SU( d )$;
	\item $M \in \mathcal{C}_{d}^{1}(2) / \mathcal{C}_{d}^{1} (1) $;
\end{enumerate}
\end{defin}
We outline the motivation for these criteria and remark that $M$ can be remembered as short for \textit{magic}.  Conditions 1-3 will be directly related to the transversality of the gate for our quantum Reed-Muller codes.   Furthermore, if we express the eigenvalues of $M$ as  $\exp(i 2 \lambda_{j} \pi / d^{m})$ then condition 2 entails $\lambda_{j}$ are integers and condition 3 is satisfied when $\sum_{j} \lambda_{j}=0$.  Condition 4 requires that while $M$ is a member of the second level of the Clifford hierarchy, it is not a member of the Clifford group itself.  From this we conclude that the operator
\begin{equation}
	C_{M} = M X M^{\dagger}
\end{equation}
is in the Clifford group but is not a Pauli operator.  The eigenstates of $C_{M}$ will be the attractor of our distillation protocols, which is why it is essential that $C_{M}$ is a non-Pauli operator.  Distillation would be possible without requiring that $C_{M}$ is a Clifford operator, but demanding this property provides us with tools that improve the protocols efficiency.  We observe that these sets form their own hierarchy, such that for any $m < m'$ we have $\mathcal{M}_{d}^{m} \subset \mathcal{M}_{d}^{m'}$.  This holds because, almost trivially, $M^{d^{m+1}}=(M^{d^{m}})^{d}=\id^{d}=\id$.  We remark also that if $M \in \mathcal{M}_{d}^{m}$ then $M^{\dagger} \in  \mathcal{M}_{d}^{m}$ and we use this feature throughout. 


For every such set that is non-empty we will design protocols that distill eigenstates of $C_{M}$.  However, we need to know that such gates exist.  In the qubit setting, the $\pi/8$-phase gate provides such a unitary for $m=4$.  However, for $m<4$ it is easy to check that all qubit gates with the form required by conditions (1-3) of the above definition are Clifford unitaries and so fail condition (4).  Remarkably, for all odd prime dimensions $d\ge3$ we can find such gates for $m=2$, and when $d\ge 5$ these gates exist for $m=1$~\cite{Howard12}.  Using tall brackets to denote binomial coefficients we have the following.
\begin{theorem}  
\label{thm_Mgate}
For all odd primes $d$, there exists a gate $M$ such that
\begin{enumerate}
	\item for $d=3$ we have $M \in \mathcal{M}_{d}^{m}$ for all $m \geq 2$;
	\item for prime $d \geq 5 $ we have $M \in \mathcal{M}_{d}^{m}$ for all $m \geq 1$.
\end{enumerate}
One such gate is the following
\begin{equation}
	M = \sum_{j} \exp( i 2 \lambda_{j} \pi / d^{m} ) \kb{j}{j} ,
\end{equation}
with
\begin{equation}
	\lambda_{j} = d^{m-2}\biggl(d \binom{j}{3}-j\binom{d}{3}+\binom{d+1}{4}\biggr) .
\end{equation}
We refer to this $M$ as the canonical $\mathcal{M}_{d}$ gate.
\end{theorem}
In particular, the canonical $\mathcal{M}_{d}$ gate is associated with the non-Pauli Clifford unitary
\begin{equation}
	C_{M} = MXM^{\dagger}  \propto   X P ,
\end{equation}
where $P$ is the Clifford gate introduced earlier in Eq.~(\ref{eqn_Cliffords}). Clearly, a different $M$ exists for every dimension $d$. For notational clarity we suppress this $d$ dependence. The proof is elementary, but for completeness given in App.~\ref{QgateApp}.  That odd prime dimensions can produce the desired gates with smaller $m$ is no mere technicality, it has far reaching benefits for the magic state distillation in higher dimensions.  We also remark that these gates, for $d=3,5$, are Clifford equivalent to those found in Ref. \cite{Howard11} to be the most robust to depolarizing noise before becoming stabilizer operations.  This family of gates was discovered independently and concurrently to this paper by Howard and Vala \cite{Howard12}, and many other important properties of such gates can be found there.

The eigenstates of $C_{M}$ are non-stabilizer states, which we label $\ket{M_{k}}$. We note that $\ket{M_{k}}= M \ket{+_{k}}$, where $ \ket{+_{k}}$ is an eigenstate of $X$ with eigenvalue $\omega^{k}$.  We aim to use magic state distillation to purify copies of $\ket{M_{0}}$ from noisy copies, and in turn to use these for fault-tolerant state-injection of the magic unitary $M$.  This brings us to our main result.

\begin{theorem}
\label{Gen:distillation}
Consider any $M \in \mathcal{M}_{d}^{m}$ for any odd prime $d$ and any integer $m \geq 2$, or any odd prime $d\ge5$ and $m\geq 1$.  There exists a stabilizer operation, $\mathcal{E}$, that iteratively distils the magic state $\ket{M_{0}}$.  The map $\mathcal{E}$ takes $n=d^{m}-1$ copies of a qudit state $\rho$, where
\begin{equation}
\label{epsilon}
		\epsilon = 1-\bra{M_{0}} \rho \ket{M_{0}} .
	\end{equation}	
With non-zero probability the protocol outputs a state $\rho' \propto \mathcal{E}(\rho^{\otimes n})$ such that
	\begin{equation}
		 \epsilon' = 1-\bra{M_{0}^{\dagger}} \rho' \ket{M_{0}^{\dagger}} .
	\end{equation}
There exists a $K>0$ such that for all $\epsilon$ we have $\epsilon' \leq K \epsilon^{2}$.    Consequently, there exists a threshold $\epsilon^*>0$ such that if $0< \epsilon < \epsilon^*$ then $\epsilon' < \epsilon$.
\end{theorem}
Notice that after a single iterate, using as input noisy $\ket{M_{0}}$ states, the protocol will output a noisy $\ket{M^{\dagger}_{0}}$ state.  By performing an even number of iterations a fixed state can be distilled.  We call this phenomena cycling, and in many cases it may be prevented by some Clifford unitary correction.  However, cycling can be desirable as it provides us with a mechanism for producing both $\ket{M_{0}}$ and  $\ket{M_{0}^{\dagger}}$ states.  The rate of error suppression is always quadratic, and so these results give the first better than linear error reductions in higher dimensional systems.


The Clifford unitary $C_{M}$ plays a practical role in several steps of our protocols. First, it is used for $C_{M}-$\textit{twirling}, which is a process for converting input states into a canonical form.  By randomly choosing an integer, $k=1,..., d$ and applying $C_{M}^{k}$ we twirl any quantum state into the $\ket{M_{k}}$ basis.  Hence, all qudit states, $\rho$, can be twirled into a form that depends only on $d-1$ independent parameters, such that
\begin{equation}
\frac{1}{d}	\sum_{k \in \mathbb{F}_{d}} C_{M}^{k} \rho ( C_{M}^{k})^{\dagger} = \sum_{k} f_{k} \kb{M_{k}}{M_{k}}.
\end{equation}
Our distillation protocols seek to increase the value of $f_{0}$. We will see that $C_{M}$ is also used in our protocols for \textit{Clifford correction}, which significantly increases the success probability, and as part of the final state-injection.


\section{Magic state distillation protocols}

\subsection{CSS codes}

Calderbank, Shor and Steane identified a special class of quantum codes, which in their honor are now known as CSS codes~\cite{NC01b}.  These codes have stabilizers generated by two subgroups,  $\mathcal{S}_{Z}$ and $\mathcal{S}_{X}$, which contain only $Z^{k}$ and $X^{k}$  terms, respectively.  Therefore, the code projector has the form $\Pi_{\mathcal{S}} = \Pi_{\mathcal{S}_{X}} \Pi_{\mathcal{S}_{Z}}$.   All CSS codes, can also be described by a pair of classical vector spaces, which correspond to $\mathcal{S}_{Z}$ and $\mathcal{S}_{X}$.  If we have a vector $\vec{u} \in \mathbb{F}_{d}^{n}$ and a single qudit operator, $U$, then we define the $n$-qudit operator
\begin{equation}
	U[\vec{u}] = \otimes_{k=1}^{n} U^{u_{k}} .
\end{equation}
The $k^{\mathrm{th}}$ element of the vector, $\vec{u}$, tells us what multiple of $U$ acts on the $k^{\mathrm{th}}$ qudit.  It follows that for every $s \in \mathcal{S}_{Z}$ we can find a $\vec{u}$ such that $s=Z[ \vec{u} ]$. In fact, $\mathcal{S}_{Z} = \{  Z[\vec{u}] ; \vec{u} \in \mathcal{L}_{Z} \}$ where $\mathcal{L}_{Z} $ is a linear vector space.  The closure of the stabilizer group under multiplication is easily seen to directly correspond to closure of $\mathcal{L}_{Z} $ under additional modulo $d$.  Similarly we can find a linear code, $\mathcal{L}_{X}$, for $\mathcal{S}_{X}$.  The whole stabilizer must be Abelian and so for all $\vec{u} \in \mathcal{L}_{X}$ and $\vec{v} \in \mathcal{L}_{Z}$ we require $\langle \vec{u}, \vec{v} \rangle = \oplus_{j} u_{j} v_{j} =0 $. Furthermore, for any code, $\mathcal{L}$, we define the dual code $\mathcal{L}^{\perp} = \{ \vec{u} ; \langle \vec{u}, \vec{v} \rangle = 0 , \forall \vec{v} \in \mathcal{L} \}$.  In terms of duality,  commutation inside the stabilizer equates to $\mathcal{L}_{X} \subset \mathcal{L}_{Z}^{\perp} $ and $\mathcal{L}_{Z} \subset \mathcal{L}_{X}^{\perp} $.  The dimensionality of the duals are related by $\mathrm{Dim}( \mathcal{L}^{\perp} ) = n - \mathrm{Dim} (\mathcal{L}) $, where $n$ is the dimension  of the vector field they inhabit, namely $\mathbb{F}_{d}^{n}$.  For a CSS code $k=n-\mathrm{Dim}(\mathcal{L}_{Z})-\mathrm{Dim}(\mathcal{L}_{X})$ gives the number of logical qudits supported by the code $\Pi$.

Here we are solely interested in stabilizer codes of only $d$ dimensions, that is a single logical qudit.  It is useful to specify a basis spanning the code, which we again do using Pauli operators $Z_{L}$ and $X_{L}$.   These are the so-called logical operators of the subspace and they must commute with the code stabilizer.  Whereas, with respect to each other the logical operators must conjugate in the same way as $Z$ and $X$, such that $X_{L} Z_{L}= \omega^{-1} Z_{L} X_{L}$. It follows that there exists an orthonormal basis, $\{ \ket{j_{L}} \}$, of stabilizer states that obey $Z_{L}\ket{j_{L}}= \omega^{j}\ket{j_{L}}$, $X_{L}\ket{j_{L}}= \ket{j_{L} \oplus 1}$, and which we call the logical basis.  In this basis, the code projector can be expressed as $\Pi = \sum_{j} \kb{j_{L}}{j_{L}}$.  We also make use of the $X$-basis that we denote $\ket{+_{j}}$ for single qudits stabilized by $\omega^{-j}X$ and $\ket{+_{j}^{L}}$ for logical encoded states stabilized by $\omega^{-j}X_{L}$. Typically, such logical operators can also be expressed in terms of vectors, such as $X_{L}= X[\vec{u}]$ where commutation of $X_{L}$ with $\mathcal{S}_{Z}$ entails $\vec{u} \subset \mathcal{L}_{Z}^{\perp}$ and  $\mathcal{L}_{Z} \subset \vec{u}^{\perp}$.  

Given this vector description, a useful fact is that $\mathcal{L}_{Z} = ( \mathrm{span}( \mathcal{L}_{X} , \vec{u} ))^{\perp}$ where the $\mathrm{span}(..., ... )$ is the vector space generated by its arguments.  Let us prove this by first observing that since $\mathcal{L}_{Z} \subset  \vec{u}^{\perp}$ and  $\mathcal{L}_{Z} \subset \mathcal{L}_{X}^{\perp}$ we have that $\mathcal{L}_{Z} \subset ( \mathrm{span}( \mathcal{L}_{X} , \vec{u} ))^{\perp}$.  That $\mathcal{L}_{Z}$ can be no smaller than this set follows from dimension counting; that is 
\begin{eqnarray*}
	\mathrm{Dim}[ ( \mathrm{span}( \mathcal{L}_{X} , \vec{u} ))^{\perp} ] & = & n - \mathrm{Dim}[\mathrm{span}( \mathcal{L}_{X} , \vec{u} )]  , \\ \nonumber
	& = & n - \mathrm{Dim}(\mathcal{L}_{X}) - 1 .
\end{eqnarray*}
Since we have a single logical qudit, $k=1$, we know also that  $\mathrm{Dim}( \mathcal{L}_{Z})= n - \mathrm{Dim}(\mathcal{L}_{X})-1$.  Since the dimensionalities match, the assertion is proven.  Taking also $Z_{L}=Z[\vec{v}]$ and noting $(\mathcal{L}^{\perp})^{\perp}=\mathcal{L}$, many such results for single qudit codes can be deduced by similar reasoning, 
\begin{eqnarray} \label{ident1}
	\mathcal{L}_{Z} & = & [ \mathrm{span}( \mathcal{L}_{X} , \vec{u} )]^{\perp} , \\  \label{ident2}
	\mathcal{L}_{Z}^{\perp} & = & \mathrm{span}( \mathcal{L}_{X} , \vec{u} ) , \\ \label{ident3}
	\mathcal{L}_{X} & = & [ \mathrm{span}( \mathcal{L}_{Z} , \vec{v} )]^{\perp} , \\  \label{ident4}
	\mathcal{L}_{X}^{\perp} & = &  \mathrm{span}( \mathcal{L}_{Z} , \vec{v} ) .  
\end{eqnarray}
We employ the above relations throughout. 

The smallest unitary capable of non-trivially acting on the code gives the robustness of the code to noise.  For CCS codes it suffices to consider phase and bit flip noise separately.  For an operator $U[\vec{u}]$ its ``size" is measured by the Hamming weight, $|\vec{u}|_{H} = \{ \# x_{j} ; x_{j} \neq 0 \}$, so the number qudits upon which the operator acts non-trivially.   The robustness to phase noise is measured by the distance, $D_{Z} = \min \{ |\vec{v}|_{H} ;  Z[\vec{v}] \Pi =  Z_{L}\Pi \}$, and for bit flip noise $D_{X} = \min \{ |\vec{v}|_{H} ;  X[\vec{v}] \Pi =  X_{L}\Pi \}$.  The overall distance of the code is $D=\min \{ D_{X}, D_{Z} \}$. Finally we remark that for any code there always exists a Clifford unitary that \textit{decodes}, such that $U Z_{L} U^{\dagger} = Z_{1} $ and $U X_{L} U^{\dagger} = X_{1} $.  

\subsection{Suitable codes}

We now define the broad class of quantum codes that we show can be used to distill these magic states.
\begin{defin}
\label{Def_S_distill_code}
	An $n$-qudit stabilizer code, $\Pi$, is an $\mathcal{M}_{d}^{m}$-distillation code if all of the following hold
	\begin{enumerate}
		\item all $M \in \mathcal{M}_{d}^{m}$ are transversal such that $M^{\otimes n} \Pi (M^{\otimes n})^{\dagger}  = M_{L}^{\dagger} \Pi M_{L} $ ;
		\item it has distance, $D \geq 2$;
		\item it has logical Pauli operators $X_{L}=X[\vec{1}]$ and $Z_{L}=Z[(d-1)\vec{1}]$.
	\end{enumerate}
\end{defin}
We have introduced the vector shorthand $\vec{1}=(1,1,...1)$.  Notice that we require a special kind of transversality, such that the logical operator, $M_{L}^{\dagger}$, is implemented by applying $M^{\otimes n}$.  The need for complex transposition will be explained later, and will be seen to result in a cycling phenomenon in the distillation protocol.  

Here we show that all $\mathcal{M}_{d}^{m}$--distillation codes can be used to perform distillation for magic states of the form $\ket{M_{0}}=M\ket{+_{0}}$ for all $M \in \mathcal{M}_{d}^{m}$.  Due to cycling, after a single iterate using as input noisy $\ket{M_{0}}$ states, the protocol will output a noisy $\ket{M^{\dagger}_{0}}$ state.  
\begin{theorem}
\label{ABS:distillation}
	Given an $n$-qudit $\mathcal{M}_{d}^{m}$-distillation code of distance $D$ the following holds.  For all $M \in \mathcal{M}_{d}^{m}$ there exists a stabilizer operation, $\mathcal{E}$, that iteratively distils the magic state $\ket{M_{0}}$. The protocol takes as input $n$ copies of a state, $\rho$, where
	\begin{equation}
		\epsilon = 1-\bra{M_{0}} \rho \ket{M_{0}} .
	\end{equation}	
	With non-zero probability the protocol outputs a state $\rho' \propto \mathcal{E} ( \rho^{\otimes n})$ such that
	\begin{equation}
		 \epsilon' = 1-\bra{M_{0}^{\dagger}} \rho' \ket{M_{0}^{\dagger}}   .
	\end{equation}
There exists a $K>0$ such that for all $\epsilon$ we have $\epsilon' \leq K \epsilon^{D}$.  Consequently, there exists a threshold $\epsilon^*>0$ such that if $0< \epsilon < \epsilon^*$ then $\epsilon' < \epsilon$.
\end{theorem}
Later we show the existence of the required codes with $d=2$, which will then entail Thm.~\ref{Gen:distillation}.  For now we show how to proceed given such a code.

\subsection{The protocol}

We prove the above key result constructively. Given an $n$-qudit $\mathcal{M}_{d}^{m}$-distillation code and any $M \in \mathcal{M}_{d}^{m}$, we can perform the following iterative magic state distillation protocol.
\begin{enumerate}
	\item Take $n$ copies of the state $\rho$ and $C_{M}$-twirl;
	\item Measure generators of the phase stabilizer $\mathcal{S}_{Z}$;
	\item Accept all outcomes, but perform a Clifford correction operator $C_{M}[\vec{w}]$ tuned to outcomes;
	\item Measure generators of the bit-flip stabilizer $\mathcal{S}_{X}$;
	\item Postselect on all ``+1" measurement outcomes;
	\item Decode the encoded qudit to a single qudit;
	\item Use the output labelled $\rho'$ as input in the next iterate.
\end{enumerate}
When iterating the protocol, on the odd iterates we must replace $C_{M}$ by $C^{\dagger}_{M}$ to account for cycling.  We have not yet defined the exact setting of $C_{M}[\vec{w}]$,  but will come to this in due time.  For simplicity though, we begin with assuming that step 2 generates all ``+1" measurement outcomes, for which $C_{M}[\vec{w}]= \id$.  We explain later how the Clifford correction in step 3 increases the success probability.

After $C_{M}$-twirling the $n$-copies we have a state
\begin{equation}
	\rho^{\otimes n} = \sum_{ \vec{v} \in \mathbb{F}_{d}^{n} } \alpha_{ \vec{v}}  \kb{M_{\vec{v}}}{M_{\vec{v}}},
\end{equation}
where
\begin{equation}
	\ket{M_{\vec{v}}} = \ket{M_{v_{1}}} \ket{M_{v_{2}}}... \ket{M_{v_{n}}} ,
\end{equation}
and
\begin{equation}
	\alpha_{\vec{v}} = \prod_{k \in \mathbb{F}_{d}} f_{k}^{\wt_{k}(\vec{v})} ,
\end{equation}
where $\wt_{k}(\vec{v})$ is the $k$-weight, the number of elements in $\vec{v}$ equal to $k$, and $f_{k} = \langle M_{k} \vert \rho  \vert M_{k}\rangle$. We note that,  
\begin{equation}
	\rho^{\otimes n} = M_{L}^{\dagger} \left( \sum_{ \vec{v} \in \mathbb{F}_{d}^{n} } \alpha_{\vec{v}} \kb{+_{\vec{v}}}{+_{\vec{v}}} \right) M_{L},
\end{equation}
where $M_{L}^{\dagger}=M^{\otimes n}$.   Upon a successful projection onto the code subspace, we have
\begin{equation}
	\Pi \rho^{\otimes n}  \Pi = M_{L}^{\dagger} \left( \sum_{ \vec{v} \in \mathbb{F}_{d}^{n} } \alpha_{\vec{v}}  \Pi   \kb{+_{\vec{v}}}{+_{\vec{v}}}  \Pi \right) M_{L},
\end{equation}
as the projector commutes with $M_{L}$.  We need to determine the effect of each term $\Pi \ket{+_{\vec{v}}}$, which we will find to be
\begin{eqnarray}
\label{Eq_NoError}
	\Pi \ket{+_{\vec{v}}} & = & 0 ; \forall \vec{v} \notin \mathcal{L}_{X}^{\perp} ; \\ 
\label{Eq_Error}
	\Pi \ket{+_{\vec{v}}} & = & \sqrt{c} \ket{+^{L}_{j}}  ; \forall \vec{v} \oplus j \vec{1} = \vec{w} , \mathrm{s.t.} \vec{w} \in \mathcal{L}_{Z}.
\end{eqnarray}
The first equation covers all $ \vec{v} \notin \mathcal{L}_{X}^{\perp}$ and the second equation covers all $\vec{v} \in \mathrm{span} ( \mathcal{L}_{Z} , \id) $.  By virtue of Eq.~($\ref{ident4}$) we know $ \mathcal{L}_{X}^{\perp}=\mathrm{span} ( \mathcal{L}_{Z} , \vec{1}) $ and so these equations account for all possible $\vec{v}$.  The constant $c$ gives the probability of this projection when the initial state is pure,
\begin{equation}
	c=\tr( \Pi \kb{+_{0}}{+_{0}}^{ \otimes n} ) .
\end{equation}
Furthermore, $\ket{+}^{\otimes n}$ is an eigenstate of $\Pi_{\mathcal{S}_{X}}$ and so this randomness can be completely attributed to the $Z$ stabilizer measurements, which can be made deterministic by Clifford correction.  Equations~(\ref{Eq_NoError},\ref{Eq_Error}) follow directly from properties of error correcting codes, but for completeness more details are given in App.~\ref{APP_projection}.

In summary, the transversality of $M$ allows us to consider the distillation of magic states $\ket{M_{0}}$ as equivalent to the simpler problem of distillation in the $X$ basis.  Combining these results, we have
\begin{eqnarray}
	\Pi \rho^{\otimes n} \Pi =c M_{L}^{\dagger} \left( \sum_{j\in \mathbb{F}_{d}} \sum_{ \vec{v} \oplus j \vec{1} \in \mathcal{L}_{Z} } \alpha_{\vec{v}}  \kb{+_{j}^{L} }{+_{j}^{L}}  \right) M_{L}.
\end{eqnarray}
The output state is diagonal in the basis $M^{\dagger}_{L} \ket{+^{L}_{j}}$ rather than the desired $M_{L} \ket{+^{L}_{j}}$.  We reiterate that this cycling is not problematic as an even number of iterations always brings us back to the initial basis.  Decoding onto a single qudit and using $\mathcal{E}$ to denote the whole process, we have
\begin{equation}
	\rho' \propto \mathcal{E}(\rho^{\otimes n}) =c  \sum_{j\in \mathbb{F}_{d}} \sum_{ \vec{v} \oplus j \vec{1}\in \mathcal{L}_{Z} } \alpha_{\vec{v}}     \kb{M^{\dagger}_{j} }{M^{\dagger}_{j}}  .
 \end{equation}
By expanding out $\alpha_{\vec{v}}$, we get an iterative formula for $f_{k}'= \bra{M_{k}} \rho' \ket{M_{k}}$, such that
 \begin{equation}
 \label{Eq_iterative}
 	f_{j}' =\frac{ \sum_{\vec{v}\oplus j \vec{1} \in \mathcal{L}_{Z}}  \prod_{k \in \mathbb{F}_{d}}  f_{k}^{\wt_{k} (\vec{v})} }{  P  } ,
 \end{equation}
which has been renormalized by dividing through by the success probability $P$.  This probability equals the sum of the numerators, which is
 \begin{equation}
 	P = \sum_{j \in \mathbb{F}_{d}} \sum_{\vec{v} \oplus j \vec{1} \in \mathcal{L}_{Z} }  \prod_{k \in \mathbb{F}_{d}}  f_{k}^{\wt_{k} (\vec{v})}  .
 \end{equation}
The summation over all $j$, such that $\vec{v} \oplus j \vec{1} \in \mathcal{L}_{Z}$, is equivalent to a sum over all $\vec{v} \in \mathrm{span}( \mathcal{L}_{Z}, (q-1)\vec{1})$.  Using the features of CSS codes (see Eq. \ref{ident3}) we know $\mathrm{span}( \mathcal{L}_{Z}, (d-1)\vec{1})= \mathcal{L}_{X}^{\perp}$ and so
  \begin{equation}
 	P = \sum_{\vec{v} \in \mathcal{L}_{X}^{\perp} }  \prod_{k \in \mathbb{F}_{d}}  f_{k}^{\wt_{k} (\vec{v})}  .
 \end{equation}
Notice that we have dropped a factor of $c$ from the success probability, which will be justified later by Clifford correction.  Both numerator and denominator of $f_{j}'$ are polynomials of degree $n$, and can be calculated from the classical codes.  

\subsection{Analyzing the iterative formulae}

Here we consider some properties of the above iterative formulae.  First we consider simple depolarizing noise model and give a Taylor series approximation.  Next, we consider a completely general noise model and show the existence of a distillation threshold.

When the noise is depolarizing, and so $f_{j \neq 0}=\epsilon / (d-1)$ and $f_{0}=1 - \epsilon$, the formula for the fidelity simplifies to
 \begin{equation}
 f_{0}'= \frac{\sum_{\vec{v}\in\mathcal{L}_{Z}}f_{0}^{n-|\vec{v}|_{H}}f_{j\neq 0}^{|\vec{v}|_{H}}}{\sum_{\vec{v}\in\mathcal{L}_{X}^{\perp}}f_{0}^{n-|\vec{v}|_{H}}f_{j\neq 0}^{|\vec{v}|_{H}}} ,
 \end{equation}
 where $|...|_{H}$ is again the Hamming weight.  The factors $f_{0}^{n}$ appear on both numerator and denominator and so cancel.  Making use of the shorthand
 \begin{equation}
	 \mu = \frac{f_{j \neq 0} }{f_{0}} = \frac{\epsilon}{(d-1)(1-\epsilon)} ,
\end{equation}
we can further simplify the fidelity formula to
 \begin{equation}
 \label{Eq_preMacWill}
 	f_{0}'=\frac{\sum_{\vec{v} \in \mathcal{L}_{Z}}   \mu^{|\vec{v}|_{H}} }{\sum_{\vec{v} \in \mathcal{L}_{X}^{\perp}} \mu^{|\vec{v}|_{H}}   } .
 \end{equation}
Such cases are easier to study as they depend on only a single parameter and the simple Hamming weights.  Indeed, we will show later, in Sec.~\ref{MacWill}, that this simple form can be further simplified by leveraging some powerful techniques from classical coding theory.  For now we make some casual observations concerning quadratic error suppression.  

Taylor expanding the numerator and denominator to second order we have
\begin{equation}
 	f_{0}' \sim \frac{1 + a \mu^{D} + O(\mu^{D+1})}{1+ b \mu^{D} + O( \mu^{D+1})} , 
\end{equation} 
where $a$ ($b$) is the number of weight $d$ elements of $\mathcal{L}_{Z}$ ($\mathcal{L}_{X}^{\perp}$).  Both $\mathcal{L}_{Z}$ and $\mathcal{L}_{X}^{\perp}$ contain a single weight zero element, $\vec{v}=\vec{0}=(0,0...0)$. By definition both contain no other elements with weights smaller than $d$.  Further approximating the denominator and using $f'_{0}=1-\epsilon'$ yields
\begin{equation}
 	\epsilon' \sim  (b-a) \mu^{D} + O(\mu^{D+1}).
\end{equation}  
So the suppression of errors is degree $D$ as $\mu \sim \epsilon$.  In particular, since $D \geq 2$ the error suppression is at least quadratic.  

The depolarizing noise model is useful for illustrating the salient features of a distillation protocol.  However, it is important to demonstrate error suppression and existence of a threshold for all possible noise models.  Again we rescale the noise parameters to $\mu_{k} = f_{k} / f_{0}$, and so
 \begin{equation}
 	f_{0}' =\frac{ \sum_{\vec{v} \in \mathcal{L}_{Z}}   \prod_{k=1}^{d-1} \mu_{k}^{\mathrm{wt}_{k}(\vec{v}) }} {\sum_{\vec{v} \in \mathcal{L}_{X}^{\perp}}    \prod_{k=1}^{d-1} \mu_{k}^{ \mathrm{wt}_{k}(\vec{v})  }} .
	 \end{equation}
Both $\mathcal{L}_{Z}$ and $\mathcal{L}_{X}^{\perp}$ contain $\vec{v}=\vec{0}$ for which $\mathrm{wt}_{k \neq 0}(\vec{v})=0$, and so both numerator and denominator contain a term equal to 1.  We make a very coarse lower bound on the numerator, which must be greater than 1 since all terms are positive.  We wish to upper bound the denominator less coarsely.  First we define $\mu= \max_{k \neq 0} \{ \mu_{k} \}$ and use it to replace all other noise parameters in the denominator, yielding the inequality
 \begin{equation}
 	f_{0}' \geq \left( \sum_{\vec{v}  \in \mathcal{L}_{X}^{\perp} }  \mu^{ | \vec{v} |_{H}  } \right)^{-1}.
 \end{equation}
Recall that $\vec{v}=\vec{0}$ contributes 1 to the summation and all other terms are upper bounded by $\mu^{D}$ where $D$ is the distance of the code.  Hence we have
 \begin{equation}
 	f_{0}' \geq \left( 1 + C \mu^{D} \right)^{-1} ,
 \end{equation}
 where $C$ is the number of nontrivial terms, $C=| \mathcal{L}_{X}^{\perp}|-1$.  Clearly for real $x$ we have $1 \geq (1-x^{2})$ and so $1 \geq (1-x)(1+x)$ and positivity of $x$ entails $(1+x)^{-1} \geq (1-x)$. Using this with $x=C \mu^{D}$ gives $f'_{0} \geq 1- C \mu^{D}$ and furthermore
  \begin{equation}
 	\epsilon' \leq  C \mu^{D} \leq C \left( \frac{\epsilon}{1-\epsilon} \right)^{D}  ,
 \end{equation} 
 where $\epsilon'=1-f_{0}'$.   We assume without loss of generality that $f_{0}$ is larger than all other $f_{j}$, which allows us to bound $(1-\epsilon)^{-1} \leq d$ and so
  \begin{equation}
 	\epsilon'  \leq d^{D} C \epsilon^{D} .
 \end{equation}  
This gives us a valid constant $K = d^{D}C$ as asserted in Thm. \ref{Gen:distillation} and Thm. \ref{ABS:distillation}.   The existence of some distillation threshold follows quickly.  If we consider $\epsilon^* = K^{-(D-1)^{-1}}$, we find that if $0<\epsilon< \epsilon^*$ then $\epsilon' < \epsilon$.   The above analysis is very general, but the corresponding bounds will be far from tight and a much higher $\epsilon^{*}$ will exist.
  
\subsection{Clifford correction}  
  
So far we have assumed that the $Z$ stabilizer measurements all yield the desired $``+1"$ outcome. Next we consider the process of Clifford correction, as outlined by step 3 of our protocol.  This additional strategy significantly increases the success probability of each round, so much so that success is guaranteed in the limit of pure initial states.  The general idea is that for any measurement outcomes, with resulting projector $\Pi_{\mathcal{S}_{Z}}'$, there exists a Clifford $C_{M}[\vec{w}]$ such that $C_{M}[\vec{w}]\Pi_{\mathcal{S}_{Z}}' =\Pi_{\mathcal{S}_{Z}}C_{M}[\vec{w}]$.  The key fact exploited is that for a single qudit $C_{M} Z = \omega^{-1} Z C_{M}$, and so for many qudits $C_{M}[\vec{w}]Z[\vec{v}]=\omega^{- \langle \vec{w} , \vec{v} \rangle }Z[\vec{v}]C_{M}[\vec{w}]$.  To proceed we must specify the projector $\Pi'_{\mathcal{S}_{Z}}$.  We begin by expressing the linear code as $\mathcal{L}_{Z} = \{ G\vec{u}  : \vec{u} \in \mathbb{F}_{d}^{m}  \}$ where $m=\mathrm{Dim}(\mathcal{L}_{Z})$ and $G$ is an $m$ by $n$ matrix called the generator matrix of $\mathcal{L}_{Z}$.   Each column of $G$ gives an individual generator of $\mathcal{L}_{Z}$ and hence $\mathcal{S}_{Z}$.  When the measurement corresponding to the $j^{\mathrm{th}}$ generator gives outcome $\omega^{k_{j}}$, the resulting projection is
\begin{equation}
	\Pi_{\mathcal{S}_{Z}}' =\frac{1}{2^{m}} \sum_{\vec{u} \in \mathbb{F}_{d}^{m}} \omega^{\langle \vec{k}, \vec{u} \rangle} Z[ G\vec{u}]  .
\end{equation}
Conjugating with a Clifford correction $C_{M}[\vec{w}]$ yields
\begin{equation}
	C_{M}[\vec{w}]	\Pi_{\mathcal{S}_{Z}}' =\frac{1}{2^{m}} \sum_{\vec{u} \in F_{d}^{m}} \omega^{\langle \vec{k} , \vec{u} \rangle - \langle \vec{w} , G\vec{u} \rangle } Z[ G\vec{u}] C[\vec{w}] ,
\end{equation}
and so the correction works when for all  $\vec{u}$ we have $\langle \vec{k} , \vec{u} \rangle =\langle \vec{w} , G\vec{u} \rangle$ mod $d$.  We can always choose a canonical form for the generator matrix, such that $G = ( \id_{m} | G') $, where the identity acts on the first $m$ rows of $G$ and $G'$ labels the remainder of the matrix.  For such a canonical generator matrix we choose $\vec{w}$ to equal $\vec{w} = ( k_{1},k_{2}... k_{m},0,0,...0)$ so it matches the measurement outcomes on the first $m$ entries.  This yields $\langle \vec{w}, G\vec{u} \rangle = \langle \vec{k}, \vec{u} \rangle $ and so Clifford correction achieves its goal.
 
\section{Reed-Muller codes}
 
\subsection{Some concrete examples}
 
Our demonstration of magic state distillation in higher dimensions was conditional on the existence of a $\mathcal{M}_{d}^{m}$-distillation codes, as specified in Def.~\ref{Def_S_distill_code}.  Before introducing a family of $\mathcal{M}_{d}^{m}$-distillation codes for all odd prime $d$, we give some concrete examples.  We label the codes as $\mathcal{QRM}_{d}(m)$ where $d$ is again the dimensionality and $m$ dictates the codes size and transversality properties.
\begin{defin}
\label{QRM32}
$\mathcal{QRM}_{3}(2)$ is a CSS code over $n=8$ qudits of dimension 3.  The $\mathcal{L}_{X}$ code is generated by
\begin{eqnarray}
	\vec{u}_{1} = ( 1 , 2, 0, 1, 2, 0, 1, 2) , \\ \nonumber
	\vec{u}_{2} = ( 0 , 0, 1, 1, 1, 2, 2, 2) . 
\end{eqnarray}
Whereas, $\mathcal{L}_{Z}$ is the code generated by
\begin{eqnarray}
	\vec{v}_{1} & = & ( 1,2,0,1,2,0,1,2) ,	\\ \nonumber
	\vec{v}_{2} & = & (0,0,1,1,1,2,2,2),	\\ \nonumber
	\vec{v}_{3} & = & (0,0,1,2,0,2,1,0),	\\ \nonumber
	\vec{v}_{4} & = & (1,1,0,1,1,0,1,1),	\\ \nonumber
	\vec{v}_{5} & = & (0,0,1,1,1,1,1,1).	
\end{eqnarray}
The logical operators are $Z_{L}=Z[ 2 \vec{1}]$ and $X_{L}=X[\vec{1}]$.
\end{defin}
For the above qutrit code, we will find that it is transversal with respect to the canonical $\mathcal{M}_{3}$ non-Clifford gate, as in Thm.~\ref{thm_Mgate}, 
\begin{equation}
\label{Mgate3}
	M = \left( \begin{array}{ccc}  \tau & 0 & 0 \\
							0 & 1 & 0 \\
							0 & 0 & \tau^{-1}
	\end{array} \right) ,
\end{equation}
where $\tau = \exp ( i 2 \pi / 9)$.  

In the introduction of suitable non-Clifford gates,  Thm.~\ref{thm_Mgate} showed that for odd primes greater than 3, it was sufficient to set $m=1$ to find non-Clifford gates.  Remarkably, this means that we can find even smaller codes with transversal non-Clifford gates.  The smallest such code exists for $d=5$, and as we shall see later it performs exceptionally well at magic state distillation.
\begin{defin}
\label{QRM51}
$\mathcal{QRM}_{5}(1)$ is a CSS code over $n=4$ ququints of dimension 5.  The $\mathcal{L}_{X}$ code is generated by
\begin{equation}
	\vec{u}_{1} = ( 1 , 2, 3, 4) .  
\end{equation}
Whereas, $\mathcal{L}_{Z}$ is the code generated by
\begin{eqnarray}
	\vec{v}_{1} & = &  ( 1 , 2, 3, 4),	\\ \nonumber
	\vec{v}_{2} & = & (1 , 4, 4, 1),	
\end{eqnarray}
The logical operators are $Z_{L}=Z[ 4 \vec{1}]$ and $X_{L}=X[\vec{1}]$.
\end{defin}
For the above code we find it is transversal with respect to the canonical $\mathcal{M}_{5}$ non-Clifford gate, 
\begin{equation}
\label{Mgate5}
	M = \left( \begin{array}{ccccc}  \omega^3 & 0 & 0 & 0 & 0 \\
							0 & \omega & 0 & 0 & 0 \\
							0 & 0 & \omega^{-1} & 0 & 0 \\
							0 & 0 & 0 & \omega^{-2} & 0 \\
							0 & 0 & 0 & 0 & \omega^{-1} 
	\end{array} \right) ,
\end{equation}
where  here $\omega = \exp ( i 2 \pi / d) = \exp ( i 2 \pi / 5)$.  Notice how the eigenvalues are all multiples of $\omega$.  For dimensions smaller than $d=5$ any diagonal gate with phases that are multiples of  $\omega$ will be a Clifford gate rather than a non-Clifford as desired.  This property makes it possible in higher dimensions to find smaller codes with a transversal non-Clifford.

From this information one can numerically verify that both codes are well defined and have the correct transversality properties. Transversality can be verified by calculating the effect of the non-Clifford gates on the logical basis states. Over the following sections we develop an analytic proof that these features are valid for a whole family of quantum codes.  Further details of the performance of these codes are given later, but we hope the examples help guide the reader through the general case.



\subsection{Classical Reed-Muller codes}

Here we review $d$-ary generalizations of Reed-Muller codes~\cite{RM68,RM68b,RM70,RM98} and derive the crucial properties we exploit later.  Convention dictates that we denote Reed-Muller codes as $\mathcal{RM}_{d}(u,m)$, where $d$ tells us the relevant field, $u$ is the order of code and $m$ determines the size of the code.  Here we explicitly use only Reed-Muller codes of 1$^{\mathrm{st}}$ order, so $u=1$.  All Reed-Muller codes are defined by polynomials of a degree bounded by the codes order.   For order 1 Reed-Muller codes we must consider degree 1 polynomials, that is linear functions. The dual of a Reed-Muller is another Reed-Muller code, though it may have a different order ~\cite{RM68,RM68b,RM70,RM98}.  In this way higher order Reed-Muller codes do enter into our work.  However, it is sufficient for us to define them in terms of duality.  Ultimately, we will not use these codes but their smaller shortened versions introduced in the next section.  However, for pedagogical reasons we first review the unshortened variants.

We begin with a review of linear maps.  There are $d^{m}$ linear maps from $\mathbb{F}_{d}^{m}$ onto $\mathbb{F}_{d}$.  All such maps,  $g_{\vec{\bar{u}}}: \mathbb{F}_{d}^{m} \rightarrow \mathbb{F}_{d}$,  can be labeled by vectors themselves, say $\vec{\bar{u}} \in \mathbb{F}_{d}^{m}$, and then the function will evaluate to $g_{\vec{\bar{u}}}(\vec{a})=\langle \bar{\vec{u}} , \vec{a} \rangle =\oplus_{j}\bar{u}_{j}a_{j}$, again modulo $d$.   Next, we consider another mapping, $U_{d}^{m}: \mathbb{F}_{d}^{m} \rightarrow \mathbb{F}_{d}^{n}$, where $n=d^{m}$, such that
\begin{equation}
	 U_{d}^{m}(\vec{\bar{u}})= (  \langle \vec{\bar{u}} , \vec{a}_{0} \rangle , \langle \vec{\bar{u}} , \vec{a}_{1} \rangle ,... \langle \vec{\bar{u}} , \vec{a}_{n-1} \rangle   )  ,
\end{equation}
where $\vec{a}_{j}$ is the base $d$ representation of the natural number $j$.  For example, with $d=3$ and $m=2$ we would have the ordered set
\begin{eqnarray*}
	\{ \vec{a}_{j} \}& = & \{ (0,0),(0,1), (0,2), (1,0), \\ 
	& & (1,1), (1,2), (2,0),(2,1),(2,2) \}  .
\end{eqnarray*}
Hence for $\vec{\bar{u}}=(0,1)$ we have
\begin{equation}
\label{unit}
	U^{2}_{3} [ \vec{\bar{u}} ]= U^{2}_{3} [ (0,1) ] = ( 0, 1 , 2, 0 , 1 , 2 , 0, 1 , 2 ).
\end{equation}
For any $d$ and $m$ (positive integers), the set $\mathcal{L} = \{ \vec{u} = U_{d}^{m}(\vec{\bar{u}}) ; \vec{\bar{u}} \in \mathbb{F}_{d}^{m}  \}$ is a linear vector space.  Closure of the vector space under addition follows directly from the closure under addition of homogenous linear maps.  The codes of interest are constructed by considering all affine functions, which are linear maps plus an additional constant $c$ such that they map $\bar{\vec{u}}$ to $U_{d}^{m}(\vec{\bar{u}}) \oplus c\vec{1}$.
\begin{defin}
	Unshortened Reed-Muller codes, $\mathcal{RM}_{d}(1,m)$, are classical linear codes on $\mathbb{F}_{d}^{n}$, where $n=d^{m}$, of dimension $m+1$.  They are the set of codewords $\mathcal{RM}_{d}(1,m)= \{  U_{d}^{m}(\vec{\bar{u}}) \oplus c\vec{1} : \vec{\bar{u}} \in \mathbb{F}_{d}^{m}, c \in \mathbb{F}_{d}  \}$ defined in terms of affine functions.
\end{defin}
Such codes have many exotic properties.  Before investigating them we introduce one more definition.
\begin{defin}
\label{Lfunction}
	We say a function $\Lambda: \mathbb{F}_{d}^{n} \rightarrow \mathbb{Z}$ is a $\lambda$-function if there exists a set of $d$ integers $\{ \lambda_{0},..\lambda_{d-1} \}$ such that $\sum_{j \in \mathbb{F}_{d}}\lambda_{j}=0$ and
\begin{equation}
	\Lambda(\vec{v}) = \sum_{j=1}^{n} \lambda_{v_{j}} .
\end{equation}
\end{defin}
The reader should note that $\lambda$-functions are closely related to the non-Clifford gates introduced in Def.~\ref{Def_Mgate}.  Our main observation here is the following.
\begin{lem}
\label{unPun}
	Given a $\lambda$-function $\Lambda$ and an unshortened code $\mathcal{RM}_{d}(1,m)$ all $\vec{v}\in \mathcal{RM}_{d}(1,m)$ satisfy $\Lambda(\vec{v})=0$ mod $d^{m}$.
\end{lem}
To prove the lemma we first consider codewords where $\vec{\bar{u}}=\vec{0}$, and so $\vec{v}=(c,c,c...c)$, then 
\begin{equation}
	\Lambda(\vec{v}) = d^{m} \lambda_{c} ,
\end{equation}
which vanishes modulo $d^{m}$.  Let us now consider the codeword for the unit vector,  $\vec{\bar{u}}=(1,0,0..0)$, and $c=0$.  The corresponding codeword has a repetitive structure as in Eq. (\ref{unit}), where each element of $\mathbb{F}_{d}$ appears $d^{m-1}$ times. Hence,
\begin{eqnarray}
	\Lambda(\vec{v})  &=& d^{m-1}  \sum_{j=0}^{d-1}  \lambda_{j} = 0,  \\ \nonumber 
\end{eqnarray}
since we required in definition of a $\lambda$-function that $\sum_{j=0}^{d-1}  \lambda_{j} = 0$.  The above argument looks tailored to codewords for a unit vector $\vec{\bar{u}}$, but a similar argument holds for all codewords with non-trivial $\vec{\bar{u}}$.   That is, for any non-trivial $\vec{\bar{u}}$ there are $d^{m-1}$ different linear maps that evaluate to each possible output.  To see this, consider that the family of linear maps is invariant under change of variables that preserve linearity.  Hence, the family of functions can always be expressed in a basis such that $\vec{\bar{u}}$ \textit{is} a unit vector.  Furthermore, these codewords have uniform multiplicity of every value $\mathbb{F}_{d}$, and so adding $c \vec{1}$ will only reorder the elements and not the multiplicity with which they appear.  This proves our lemma.

In summary, unshortened Reed-Muller codes have a huge amount of symmetry that they inherit from the families of affine and linear maps.  However, they actually have too much symmetry for our purposes.  We break just enough of that symmetry by shortening the code.

\subsection{Shortened classical Reed-Muller codes}

Given a code $\mathcal{L}$ over $\mathbb{F}_{d}^{n}$, the corresponding shortened code, denoted $\mathcal{L}^{*}$, is over $\mathbb{F}_{d}^{n-1}$.  It contains all the codewords of $\mathcal{L}$ with 0 in the first position and that position deleted.   The process of shortening is closely related to puncturing, where the first position is removed but all codewords are kept.  We can also give a self contained definition of a shortened Reed-Muller code as follows. 
\begin{defin}
	Shortened Reed-Muller codes, $\mathcal{RM}^{*}_{d}(1,m)$, are classical linear codes on $\mathbb{F}_{d}^{n}$, where $n=d^{m}-1$, of dimension $m$.  They are the set of codewords $\mathcal{RM}^{*}_{d}(1,m)= \{  P_{d}^{m}(\vec{\bar{u}}) : \vec{\bar{u}} \in \mathbb{F}_{d}  \}$ defined in terms of linear maps.
\end{defin}
Here $P_{d}^{m}$ is the same map as $U_{d}^{m}$, but omitting the first element.  For example, the shortened version of Eq.~(\ref{unit}) is
\begin{equation}
  	P^{2}_{3} [ \vec{\bar{u}} ]= P^{2}_{3} [ (0,1) ] = ( 1 , 2, 0 , 1 , 2 , 0, 1 , 2 ),
\end{equation}
which the reader may have noticed is also one of the generators of the $\mathcal{L}_{X}$ code for the quantum $\mathcal{QRM}_{3}(2)$ code reviewed earlier.   Notice that the self-contained definition of the shortened Reed-Muller code makes use of only linear maps and not affine maps.  In the unshortened code we had a generator $\vec{1}$ that corresponded to the constant term in affine functions.  However, when shortening a code we only keep codewords with zero in the first position and so the $\vec{1}$ generator is dropped.  For this reason the dimension of the code drops by one;  $\Dim(  \mathcal{RM}^{*}_{d}(1,m)  )=\Dim(  \mathcal{RM}_{d}(1,m)  ) -1$. Let us now consider the shortened analog of Lem.~\ref{unPun}.
\begin{lem}
\label{Pun}
	Given a $\lambda$-function $\Lambda$ and a shortened code $\mathcal{RM}^{*}_{d}(1,m)$ all $\vec{v} \in \mathcal{RM}^{*}_{d}(1,m)$ satisfy $\Lambda(\vec{v} \oplus c \vec{1})=-\lambda_{c}$ mod $d^{m}$.
\end{lem}
This follows quickly from Lem.~$\ref{unPun}$.  Given a $\vec{v} \in \mathcal{RM}^{*}_{d}(1,m)$, let us define
\begin{eqnarray}
	\vec{w}&=&(0, v_{1}, v_{2},.. v_{n}) \oplus c \vec{1} , \nonumber \\ 
	&=&(c, v_{1} \oplus c, v_{2} \oplus c,... v_{n} \oplus n) ,
\end{eqnarray}
where clearly $\vec{w}$ is a codeword of the unshortened code $\mathcal{RM}_{d}(1,m)$. Furthermore, $\Lambda(\vec{w})=\Lambda(\vec{v}) + \lambda_{c}$ as it has an extra term appended.  However, Lem.~$\ref{unPun}$ tells us that $\Lambda(\vec{w})=0$ and so $\Lambda(\vec{v}) = - \lambda_{c}$.  We will soon see that Lem \ref{Pun} is intimately related to transversality of quantum gates for an associated quantum code.


\subsection{Quantum Reed-Muller codes}

Here we construct quantum codes from shortened Reed-Muller codes for general $m$ and $d$.
\begin{defin}
$\mathcal{QRM}_{d}(m)$ with $m\geq 1$ is a quantum CSS code over $n=d^{m}-1$ qudits of prime dimension $d$.  The codespace is defined by
\begin{enumerate}
 	\item  $\mathcal{L}_{X}=\mathcal{RM}^{*}_{d}(1,m)$;
	\item  $\mathcal{L}_{Z} = [ \mathrm{span}( \mathcal{L}_{X} , \vec{1} ) ]^{\perp}$;
	\item  $X_{L}=X[\vec{1}]$;
	\item  $Z_{L}=Z[(d-1)\vec{1}]$.
\end{enumerate}
\end{defin}
We could have equivalently specified $\mathcal{L}_{Z}$ as a higher order Reed-Muller code,  though the above is simpler.  We first check that $\mathcal{QRM}_{d}(m)$ codes are indeed quantum codes.  By construction, the stabilizer is Abelian as $\mathcal{L}_{Z} \subset \mathcal{L}_{X}^{\perp}$.   It is easy to check the logical operators are well defined: that $Z_{L}$ commutes with the stabilizer; $X_{L}$ commutes with the stabilizer; and $X_{L}Z_{L}=\omega^{-1} Z_{L}X_{L}$.   Now our next main result can be concisely stated.
\begin{theorem}
$\mathcal{QRM}_{d}(m)$ quantum codes are $\mathcal{M}_{d}^{m}$-distillation codes of distance $D = 2$.
\end{theorem}
The main property we need to prove is transversality for all $M \in \mathcal{M}_{d}^{m}$.  As with all CCS codes, we have that
\begin{equation}
	\ket{j_{L}} =\frac{1}{\sqrt{|\mathcal{L}_{X}|}} \sum_{\vec{v} \in \mathcal{L}_{X}} \ket{\vec{v} \oplus  j\vec{1}}.
\end{equation}
Acting on this logical state with $M^{\otimes n}$ gives
\begin{equation}
	M^{\otimes n}\ket{j_{L}} =\frac{1}{\sqrt{|\mathcal{L}_{X}|}} \sum_{\vec{v} \in \mathcal{L}_{X}} \exp \left( i \frac{2 \pi}{d^{m}} \Lambda (\vec{v} \oplus j\vec{1})  \right) \ket{\vec{v} \oplus j\vec{1}},
\end{equation}
where $\Lambda$ is a $\lambda$-function (Recall Def.~\ref{Lfunction}) using the integers $\{ \lambda_{j} \}$ associated with the eigenvalues of the unitary $M$.  Now we use our key lemma \ref{Pun} to conclude
\begin{eqnarray*}
	M^{\otimes n}\ket{j_{L}} &=&\frac{1}{\sqrt{|\mathcal{L}_{X}|}} \sum_{\vec{v} \in \mathcal{L}_{X}} \exp (- 2 i \pi \lambda_{j} / d^m) \ket{\vec{v} \oplus j\vec{1}}, \\ \nonumber
	& = & \exp (-2 i \pi \lambda_{j} / d^m) \ket{j_{L}} 	= M_{L}^{\dagger} \ket{j_{L}},
\end{eqnarray*}
and so we can identify $M^{\otimes n}$ with $M_{L}^{\dagger}$.   


Proving a distance lower bound is straightforward as distance 2 is the smallest non-trivial distance.   The relevant distance is $D_{z}$, the smallest $|\vec{v}|_{H}$ such that it produces a logical error $Z[\vec{v}]\Pi=Z_{L}^{j}\Pi$.  For such an operator $\vec{v} \in \mathcal{L}_{X}^{\perp}$ but $\vec{v} \neq 0$, so the phase error commutes with the $X$ stabilizer but is non-trivial.  If such an operator existed with Hamming weight 1, it would entail that there existed a qudit upon which $\mathcal{L}_{X}$ acted trivially, which there is not.  That the distance is not greater than 2 follows from the following section.

\subsection{MacWilliams identities}
\label{MacWill}

We have introduced higher dimensional Reed-Muller codes and shown that they have suitable transversality properties for magic state distllation.  Knowing the code stabilizer and using Eqs~(\ref{Eq_iterative},\ref{Eq_preMacWill}) we can calculate the exact analytic formuale for arbitrary noise.  For $\mathcal{QRM}_{3}(2)$ the general noise problem is tractable because $\mathcal{L}_{Z}$ and $\mathcal{L}_{X}^{\perp}$ are quite small sets, but the size and complexity of these sets grows rapidly with $d$ and $m$.   This is relevant because the fidelity after 1 iterate is calculated by summing over all elements in $\mathcal{L}_{Z}$ and $\mathcal{L}_{X}^{\perp}$.  By considering depolarizing noise, the problem is partially simplified by Eq.~(\ref{Eq_preMacWill}), which we restate here as
\begin{equation}
	f_{0}' = \frac{W_{\mathcal{L}_{Z}}( \mu)}{W_{\mathcal{L}_{X}^{\perp}}( \mu)} ,
\end{equation}
where $W_{\mathcal{L}}(\mu)$ is known as a weight enumerator 
\begin{equation}
	W_{\mathcal{L}}(\mu) = \sum_{v \in \mathcal{L}} \mu^{| \vec{v} |_{H} } .
\end{equation}
Weight enumerators have been extensively studied in classical coding theory~\cite{Sloane}. In particular, a weight enumerator for a code $\mathcal{L}$ can be related to the weight enumerator for the dual code $\mathcal{L}^{\perp}$ by the MacWilliams identity~\cite{Sloane}
\begin{equation}
	W_{\mathcal{L}^{\perp}}(\mu) =  d^{- \mathrm{Dim} ( \mathcal{L} )}[1+(d-1)\mu]^{n} W_{\mathcal{L}} \left( \tilde{\mu} \right)  \nonumber ,
\end{equation}
where we use the shorthand 
\begin{equation}
	\tilde{\mu}= \frac{1-\mu}{1 + (d-1)\mu}.
\end{equation}
Using $\mathcal{L}_{Z}=[\Span(\mathcal{\mathcal{L}}_{X}, \vec{1})]^{\perp}=(\mathcal{L}_{X}')^{\perp}$ (see Eq.~\ref{ident1}) and the MacWilliams identity we have
\begin{equation}
	f_{0}' = \frac{W_{\mathcal{L}_{X}'}( \tilde{\mu} )}{d W_{\mathcal{L}_{X}}( \tilde{\mu}) } .
\end{equation}
The codes $\mathcal{L}_{X}$ and $\mathcal{L}_{X}'$ are much smaller and simpler than their duals, and so the MacWilliams identities has proven extremely helpful.  Indeed, for Reed-Muller codes we can find a closed form for these enumerators. When $\mathcal{L}_{X}= \mathcal{RM}_{d}(1,m)$ we have (see App.~\ref{APPweightenum} for details)
\begin{equation}
\label{weightLX}
	W_{\mathcal{L}_{X} }(\tilde{\mu}) = 1 + (d^{m}-1) \tilde{\mu}^{(d^{m}-d^{m-1})} ,
\end{equation}
and 
\begin{eqnarray}
\label{weightLXone}
	W_{\mathcal{L}_{X}' }(\tilde{\mu}) & = & W_{\mathcal{L}_{X} }(\tilde{\mu})  \\
	& +  &   (d-1)[ \tilde{\mu}^{(d^{m}-1)}+(d^{m}-1)  \tilde{\mu}^{(d^{m}-1-d^{m-1})} ] \nonumber .
\end{eqnarray}
Combining all these formulae and reverting back to the original variables $\epsilon$ gives a closed analytic form, which is manageable albeit a bit long for reproducing here.  Rather we present the Taylor expansion to second order in $\epsilon$
\begin{equation}
\label{EqDepNoiseApprox}
	\epsilon' =  \frac{(d^{m}-1)(d-2)}{2 (d-1)} \epsilon^{2} + O[\epsilon^{3}].
\end{equation}
It is interesting that for all protocols based upon a quantum code $\mathcal{QRM}_{d}(m)$ we see quadratic error suppression for all odd prime $d$ and all $m$. Whereas the quantum Reed-Muller code used by Bravyi and Kitaev, $\mathcal{QRM}_{2}(4)$, obtained a cubic reduction, such that $\epsilon' \sim 35 \epsilon^{3}$.  Our analysis also describes the Bravyi-Kitaev protocol, the only difference being that in the qubit case we need $m\geq 4$, and so the above formula also holds for qubits.  It is intriguing to observe that the factor $(d-2)$ appears above and so the quadratic term would vanishes only in the qubit case, and so in higher dimensions these Reed-Muller codes are only distance 2.  This is one of many curious differences between qubits and odd prime dimensions.  

\section{Performance of protocols}
\label{SecPer}

Here we consider various aspects of the performance of our protocols.  We begin by showing that our protocols yield magic states at a rate that scales only polynomially with the desired final error probability.  We then use MacWilliams identities to analyse thresholds under depolarizing noise models for much larger codes.  Next we consider in more detail the performance of our protocol based on $\mathcal{QRM}_{3}(2)$ and  $\mathcal{QRM}_{5}(1)$
 
\subsection{Yields}
\label{Yields}

The overall performance of a protocol can be captured by its yield.  Given some target error probability, we calculate the yield as the expected fraction of the initial copies that achieves the goal.   By definition, for any protocol and any distillable state $\rho$, with error probability $\epsilon_{\mathrm{in}}$, there exists a number of rounds $N( \rho, \epsilon_{\mathrm{target}} )$ required to achieve $\epsilon_{\mathrm{target}}$.  If on the $k^{\mathrm{th}}$ round of distillation the success probability is $P_{k}$, the yield is simply
\begin{equation}
	Y(\rho,  \epsilon_{\mathrm{target}} ) = \prod_{k=1,.. N} \left( \frac{P_{k}}{n} \right) ,
\end{equation}
where $n$ is the again the number of copies use per iterate.  We are interested in how this scales as $\epsilon_{\mathrm{target}}$ vanishes. The success probability is continuous in $\epsilon$ and approaches 1 as $\epsilon$ vanishes, thus $P_{k}$ approaches 1 as $k$ increases. Therefore, for all $p < 1$ these exists a $c$ such that for all $k>c$ we have $P_{k}>P_{c}=p$.  This allows us to lower bound the yield such that
\begin{equation}
	Y(\rho,  \epsilon_{\mathrm{target}} ) \geq C \left( \frac{P_{c}}{n} \right)^{N-c} ,
\end{equation}
where $C$ is a constant overhead, independent of $\epsilon_{\mathrm{target}}$, that represents the yield for $c$ iterations.  Furthermore, after $c$ iterations the error probability is now $\epsilon_{c}$.  Next we observe that for a single round we know $\epsilon' \leq K \epsilon^{D}$ for some $K$, equivalently $K\epsilon' \leq (K \epsilon)^{D}$.  Therefore, the error probability after $N$ iterations, $ \epsilon_{N}$, satisfies $K \epsilon_{N} \leq (K \epsilon_{c})^{D^{N-c}} $.  Taking $K\epsilon_{c} < 1$ allows us to bound the number of iterations needed such that
\begin{equation}
	N - c < \left( \log_{2}  \left( \frac{\log(  \epsilon_{\mathrm{target}}^{-1} / K ) }{\log(\epsilon_{c}^{-1} / K  )  }  \right) \right) .
\end{equation}
For positive $a$ and $b$ we have the identity $a^{\log_{D}(b)} =b^{\log_{D}(a)} $, which combined with the above equations entails that
\begin{equation}
	Y(\rho,  \epsilon_{\mathrm{target}} ) \geq C   \left(  \frac{ \log( \epsilon_{\mathrm{target}}^{-1} / K  ) }{\log( \epsilon_{c}^{-1} / K )}  \right)^{ \log_{D} ( P_{c}  / n ) } .
\end{equation}
With the shorthand $\gamma = - \log_{D} ( P_{c}  / n ) $, which is positive, we have
\begin{equation}
	Y(\rho,  \epsilon_{\mathrm{target}} ) \geq C   \frac{\log(\epsilon_{c}^{-1} / K  )^{\gamma} }{ \log( \epsilon_{\mathrm{target}}^{-1} / K )^{\gamma}  } . 
\end{equation}
This decreases by a factor polynomial in $\epsilon_{\mathrm{target}}^{-1}$.  Conversely, the expected resource cost of distillation is the inverse yield, and this increases only polynomially in $\epsilon_{\mathrm{target}}^{-1}$.  The scaling is governed by the factor $\gamma=-\log_{D}(P_{c}/n)$, but $P_{c}$ can be taken arbitrarily close to 1.  As such, the relevant scaling parameter is $\gamma^{*}= \log_{D} ( n )$, and so
\begin{equation}
\label{yieldForm}
	Y(\rho,  \epsilon_{\mathrm{target}} ) \sim O[   \log( \epsilon_{\mathrm{target}}^{-1} / K )^{-\gamma^*}   ] . 
\end{equation}
For our protocols in odd prime dimension we will find that $D=2$ and $n=d^{m}-1$ so $\gamma^{*}= \log_{2} ( d^{m}-1 )$, which we give in table \ref{tab:yield}.

\begin{table}[h]
\centering
\begin{tabular}{|  c c c || l | l  |  l  |  l  |}
\hline
 & $d$ &  & $m=1$ &  $m=2$ &  $m=3$ & $m=4$ \\
 \hline
& 2 & &  N/A  &  N/A & N/A & 2.46497 \\
& 3 &  & N/A &  3     & 4.70044 &  6.32193 \\
& 5 & &2 & 4.58496 & 6.9542 & 9.2854 \\
& 7 & & 2.58496 & 5.58496 & 8.41785 &11.2288 \\
& 11 & &3.32193 &  6.90689 & 10.3772 & 13.8376 \\
& 13 & &3.58496 & 7.39232 & 11.1007 & 14.8017 \\
& 17 & &4 & 8.16993 & 12.2621 & 16.3498 \\
& 19 & & 4.16993 & 8.49185 & 12.7436 & 16.9917 \\
  \hline
\end{tabular}
\caption{The yield scaling parameter, $\gamma^*$,  for distillation by $\mathcal{QRM}_{d}(m)$ as governed by Eq. \ref{yieldForm}.   The smaller the value of $\gamma^*$, the more resource efficient the protocol in the limit of many iterations.    For qubit systems, the 10-to-2 protocol of Ref.~\cite{Knill12} achieves $\gamma^*= \log_{2}(5)\sim 2.32193$, which is the best known value for qubit protocols. N/A indicates not applicable, as for those parameters no non-Clifford gates exist.}
\label{tab:yield}
\end{table}

Notice that the code $\mathcal{QRM}_{5}(1)$ achieves the best yield scaling of all quantum Reed-Muller codes.  This accolade is retained by $\mathcal{QRM}_{5}(1)$ even if we compare it with all presently known magic state distillation protocols.

\subsection{Depolarizing noise thresholds}
\label{isoNoiseThresholds}

For some values of $d$ and $m$, we have used the exact expression for $\epsilon'$ to find the depolarizing noise threshold $\epsilon^{*}_{\mathrm{dep}}$ below which distillation occurs (see Table \ref{tab:thresh} below).   This should not be confused with the absolute threshold $\epsilon^{*}$ that holds for all noise models and can be smaller. The threshold gets weaker for both increasing $d$ and increasing $m$, as suggested by the above approximate formula for $\epsilon'$ (see Eq \ref{EqDepNoiseApprox}).  When we increase $m$, we increase the number of copies required per iteration but decrease the depolarizing noise threshold.  This makes it advantageous to use the smallest possible $m$ such that $M \in \mathcal{M}_{d}^{m}$.  The benefit of larger $m$ is rather that a large set of states are distilled by the protocol.    

\begin{table}[h]
\centering
\begin{tabular}{|  c c c || l | l  |  l  |  l  |}
\hline
 & $d$ & & m=1 &  $m=2$ &  $m=3$ & $m=4$ \\
 \hline
   & 2 && N/A &  N/A & N/A & 0.14148  \\
  & 3 & &N/A &		0.211001 	& 0.0657764  & 0.0214564 \\
  & 5 && 0.3631226 &       0.0614718    & 0.0119213  & 0.00236986 \\
  & 7 && 0.2322599  &       0.0291865	&  0.00409851 	& 0.000584079 \\
  & 11 && 0.1341066   &     0.0111835 	&  0.00100907	& 0.0000916717 \\
  & 13 & & 0.1106148&     0.00790156	& 0.000604487	& 0.0000464795 \\
  & 17 &&0.0818753  & 	0.00454655	& 0.000266565 	& 0.0000156773 \\
  & 19 & &0.072453 &	0.00362063  & 0.000190054   & 0.0000100014 \\
  \hline
\end{tabular}
\caption{The distillation threshold $\epsilon_{\mathrm{dep}}^{*}$ for depolarizing noise when distilled by $\mathcal{QRM}_{d}(m)$.  We include the threshold for the Brayvi-Kitaev protocol using 15-qubits, which uses a quantum Reed-Muller code $\mathcal{QRM}_{2}(4)$. N/A indicates not applicable, as for those parameters no non-Clifford gates exist.}
\label{tab:thresh}
\end{table}

If we also compare our protocols with the threshold of the BK protocol for $d=2$, the pattern of better threshold for smaller dimensions no longer holds.  We see that  the best threshold we observe is for $\mathcal{QRM}_{5}(1)$ with a fairly high threshold also observed for $\mathcal{QRM}_{3}(2)$.  There are many subtle differences in the Clifford group between odd and even dimension, and here those differences work in our favor.  In odd prime dimension we can construct smaller codes with transversal non-Clifford gates.   Our code $\mathcal{QRM}_{5}(1)$ uses 4 ququints covering a Hilbert space of dimension $5^{4}$, which to our knowledge is the smallest non-trivial stabilizer code with a transversal non-Clifford gate.   Furthermore, research to date indicates that smaller codes lend themselves to better thresholds.  A plausible explanation is that larger codes allow more undetected errors.  Most of these undetected errors will have a large Hamming weight, and so while negligible for small $\epsilon$, they will be damaging for the modest size $\epsilon$ relevant for threshold calculations.

Concerning thresholds for qubit protocols, we have focused on the comparable protocol using quantum Reed-Muller codes.  The qubit threshold can be slightly extended by using the 7-qubit Steane code~\cite{Rei01a} ($\epsilon^{*}=0.14645$) or the 5-qubit code~\cite{BraKit05} on a different class of magic states ($\epsilon^{*}=0.1719$). Though a slight improvement, both fall short of our qutrit and ququint thresholds and have much poorer yields. 

\begin{figure}
\includegraphics[width=250pt]{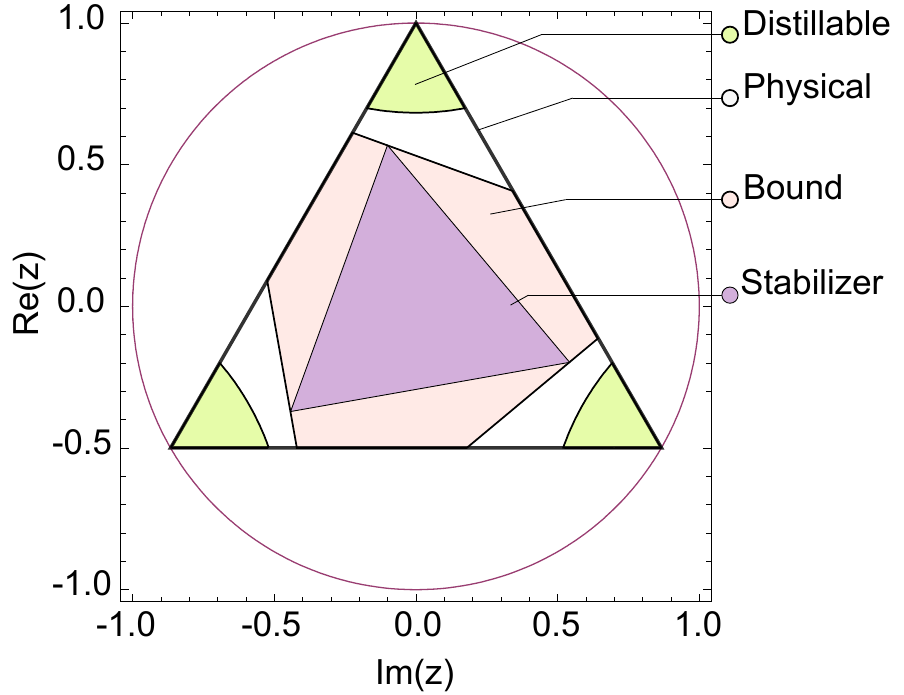}\caption{The canonical $C_{M}$-plane for a qutrit, $d=3$, which any state can be projected onto by $C_{M}$-twirling.  Every quantum state is a point in the complex plane for the complex number $z_{\rho}=\mathrm{tr}( C_{M} \rho )$.  The three pure magic states, $\ket{M_{k}}$, take values $z=1, \omega, \omega^{2}$, which have $|z|^{2}=1$ and so lie on a circle in the plane.  All \textit{physical} states have, $z = (1-f_{1}-f_{2}) + \omega f_{1}+\omega^{2}f_{2}$, and so lie in the convex hull of the pure magic states, forming a triangle of physical states.  The \textit{distillable} region of states can, by use of the $\mathcal{QRM}_{3}(2)$ protocol, be brought arbitrarily close to nearest pure magic state.  The \textit{stabilizer} states are the convex hull over the set of points, $z$, taken for each of the pure stabilizer states.  It is impossible to distil not only the stabilizer states but also the \textit{bound} states, as demonstrated in Ref.~\cite{Veitch}.  Note that the rotational symmetry is to be expected as the Pauli $Z$ rotation performs a rotation in the $C_{M}$-plane.}
\label{Cplane}
\end{figure}

Before proceeding, we will remark on our notation and terminology for quantifying depolarizing noise.  Throughout we have used $\epsilon=1-\bra{M_{0}}\rho\ket{M_{0}}$ for the error probability.  If a state suffers depolarizing noise, it has the form
\begin{equation}
	\rho = \delta \kb{M_{0}}{M_{0}} + (1-\delta) \id / d  ,
 \end{equation}
and in some parts of the literature $\delta$ is used to quantify noise.  Relating these two distinct noise measures we have
\begin{equation}
 \label{RateToProb}
	\epsilon_{\mathrm{dep}} = (d-1) \delta / d ,
 \end{equation}
and so a dependence on the dimensionality appears.  In terms of $\delta$ thresholds appear larger, with $\mathcal{QRM}_{3}(2)$ and $\mathcal{QRM}_{3}(2)$ having thresholds at $\delta=0.317$ and $\delta=0.453$ respectively.  Some readers may find using $\delta$ to be more natural as it may be related to the depolarizing noise rate of some unitary used to prepare the initial noisy magic states.  However, when unitaries suffer depolarizing noise, the best strategy is not simply to apply the noisy unitary to $\ket{+}$.  Rather better thresholds can be achieved with noisy unitaries by using the noise dilution protocol of Howard and Vala \cite{Howard12}.  Furthermore, the threshold boosts from noise dilution become more prominent for higher dimensions.


\subsection{Peformance of $\mathcal{QRM}_{3}(2)$}

\begin{figure*}
\includegraphics{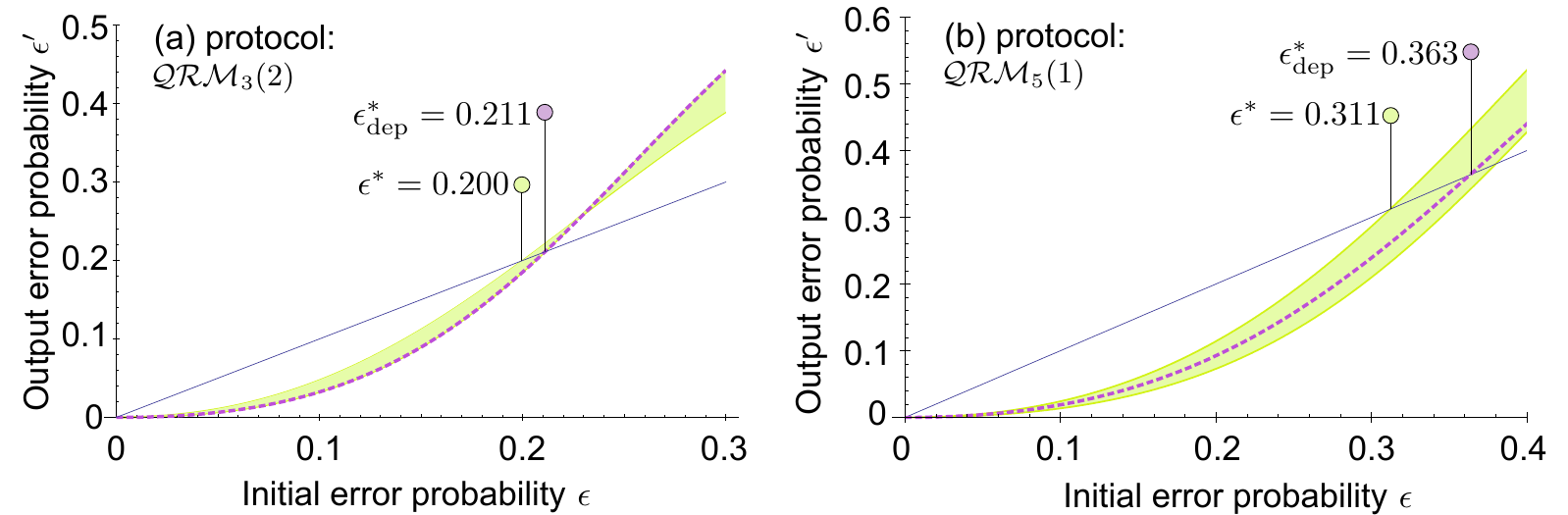}
\caption{The output error, $\epsilon'$ against input error, $\epsilon$ for (a) $\mathcal{QRM}_{3}(2)$ and (b) $\mathcal{QRM}_{5}(1)$.  For a fixed $\epsilon$ there are many different compatible states, and so there are many different possible output $\epsilon'$ and these are shown as a region rather than single curve.  For the worst case noise we mark the threshold $\epsilon^*$.  The dashed line shows the specific instance of depolarizing noise, and the associated depolarizing threshold $\epsilon^*_{\mathrm{dep}}$ is also shown.  The straight line is simple the "break even" line.}
\label{FidCombinedPlot}
\end{figure*}


Here we apply our methods to the 3 dimensional case using $\mathcal{QRM}_{3}(2)$, as explicitly defined in Def \ref{QRM32}.  In previous work~\cite{Hussain1} we have proposed other protocols for the 3-dimensional case, including a generalization of the 5-qubit code to qutrits.  While magic state distillation was observed for this 5-qutrit code, these previous studies only showed a linear suppression of noise, whereas here we observe a more rapid quadratic suppression with each iteration.  

We take $M$  to be the canonical $\mathcal{M}_{3}$ gate, as in Thm.~\ref{thm_Mgate} and Eq \ref{Mgate3}.  By $C_{M}$-twirling all single qudit quantum states are projected onto the diagonal in the $\ket{M_{k}}$ basis, such that $\rho=\sum_{k} f_{k} \kb{M_{k}}{M_{k}}$.  When we wish to distil $\ket{M_{0}}$, the weights $f_{1}$ and $f_{2}$ represent different types of noise.  A more convenient parameterization is $f_{1}=\epsilon \cos^{2}(\theta)$ and $f_{2}= \epsilon \sin^{2}(\theta)$ as we are mainly interested in how the total noise reduces.  Our techniques allow us to find an analytic solution for $\epsilon'$ after a single iterate of magic state distillation with $\mathcal{QRM}_{3}(2)$.  However, the expression is lengthy so here we truncate to 3rd order
\begin{equation}
	\epsilon'  = \epsilon^{2}[ 3 + \cos (4 \theta) ] + \epsilon^{3} [9- \cos ( 4 \theta ) ] + O[\epsilon^{4}] ,
\end{equation}
which is quadratically reduced.  In Fig (\ref{FidCombinedPlot}a), we show the exact output error probability for the whole range of different noise models (different $\theta$) and depolarizing noise ($ \theta = \pi /2`$).  We find that a threshold of $\epsilon^{*}=0.20015$ for general noise and $\epsilon^{*}_{\mathrm{dep}}=0.211001$ for depolarizing noise (as cited earlier) .  As such, for all $\theta$, if $0<\epsilon< \epsilon^{*}$ it follows that $\epsilon' < \epsilon$.   We can also find a quadratic upper bound, such that for all $\epsilon$ and $\theta$ we have  $\epsilon' \leq K \epsilon^{2}$ with $K=5.03$.  The value of $K$ is found by considering the function $\epsilon'  \epsilon^{-2}$ and numerically maximizing, so $K = \sup_{\epsilon, \theta} \{ \epsilon'  \epsilon^{-2} \}$.

The region of distillable states is actually slightly larger than the $\epsilon < \epsilon^{*}$ region, with a greater noise tolerance for some values of $\theta$.  To find the whole distillable region we resort to numerics and present the results as part of Fig.~\ref{Cplane}. Several other important regions of the plane are also highlighted.  We show the stabilizer states and bound magic states, which cannot be distilled by any stabilizer operation.  Between these regions is a non-empty regime of ambiguous status, which neither our protocol works upon nor is ruled out from distillability by any known theorem.  Even in the simple qubit case, such puzzling regimes exist and it has proven challenging to conclusively decide their status, see for example Refs.~\cite{Camp10a,Camp09c}.   

Also important is the success probability of distillation with $\mathcal{QRM}_{3}(2)$, which for all states satisfies $P \geq 1/9$ and for small $\epsilon$ is approximately
\begin{equation}
	P = 1 - 8 \epsilon + [31 + \cos ( 4 \theta ) ] \epsilon^2 + O(\epsilon^{3}).
\end{equation}
Given these fairly high success probabilities and that we use only $8$ copies per iteration, this protocol is competitive in comparison to the Bravyi-Kitaev protocol (herein BK) that also used Reed-Muller codes.  Their protocol uses 15 copies per iteration and has $P \geq 1/ 16$ and for small $\epsilon$ it achieves $P = 1 - 15 \epsilon + O(\epsilon^{2})$. Our $\mathcal{QRM}_{3}(2)$ code requires fewer copies per iteration, but it would require more iterations to achieve the same error suppression as BK, since BK has a cubic error suppression rather than just quadratic.

In Figs.~(\ref{FidYield}.1a,\ref{FidYield}.1b)  we consider the exact yield of our protocol, compared against BK, assuming depolarizing noise, so $\theta= \pi / 4$.  For small error probability $\epsilon_{\mathrm{in}}<0.05$, the yield of our protocol $\mathcal{QRM}_{3}(2)$ is similar to BK.  Both protocols give yields of the same order of magnitude and which protocol is superior fluctuates with variation in required iterations.  However, as the initial error probability $\epsilon_{\mathrm{in}}$ increases, the yield of $\mathcal{QRM}_{3}(2)$ exceeds that of BK by many orders of magnitude.  The dominant effect here is that the yield of BK vanishes as we approach the threshold $\epsilon^*_{BK}\sim 0.1415$, whereas our protocol can tolerate depolarization all the way upto $\epsilon^*_{\mathrm{dep}}\sim 0.211$.  


The results of Sec~\ref{Yields} also give us analytic tools for estimating yields. These show that for small $\epsilon_{\mathrm{target}}$ the yield of our protocol decreases as
\begin{equation}
	Y \sim O[ \log ( \epsilon_{\mathrm{target} }^{-1} / 5.03 )^{-\gamma^{*}} ]  ,
\end{equation}
where $\gamma^{*}=\log_{2}(8)=3$.  This can be compared with the BK protocol, which achieves a similar scaling with $\gamma^{*}=\log_{3}(15) \sim 2.46$.  We see these protocols have similar scaling properties, but BK performs slightly better in the large $\epsilon_{\mathrm{target} }^{-1}$ limit.   However, the numerical results reported in the previous paragraph show that finite size effects and a superior threshold often outweight these asymptotic arguments.
 
 \subsection{Peformance of $\mathcal{QRM}_{5}(1)$}
 
Next we apply our methods to the 5-dimensional case using the code $\mathcal{QRM}_{5}(1)$, as explicitly defined in Def \ref{QRM51}.   This is the first protocol ever applied to the problem of distilling magic states in 5-dimensional systems.  The code and associated protocol have many distinguishing feature already mentioned:  it is the smallest known non-trivial code to have a transversal non-Clifford; it has the largest noise threshold against depolarizing noise ($\epsilon^*_{\mathrm{dep}}=0.363$); and it has the best known scaling in terms of expected yield (with $\gamma=2$).  All these features can be attributes to the fact that $d=5$ is the smallest dimension where diagonal non-Clifford gate exist with period $d$, allowing us to work with $m=1$.

Again we take $M$  to be the canonical $\mathcal{M}_{5}$ gate, as in Thm.~\ref{thm_Mgate} and Eq.~(\ref{Mgate5}).  The $C_{M}$-twirled states are parameterised by a fidelity, $f_{0}=1-\epsilon$, and 4 independent noise parameters $f_{j}$ for $j=1,2,3,4$.  In Fig. (\ref{FidCombinedPlot}b) we show the range of different output error rates for all different types of noise and the depolarizing noise, which have thresholds of $\epsilon^{*}=0.31195$ and $\epsilon^{*}_{\mathrm{dep}}=0.363122$.  We noted earlier that the $\mathcal{QRM}_{5}(1)$ possesses the best known protection against depolarizing noise, but also see here that its robustness against generic noise is also unrivaled.

\begin{figure*}
\includegraphics{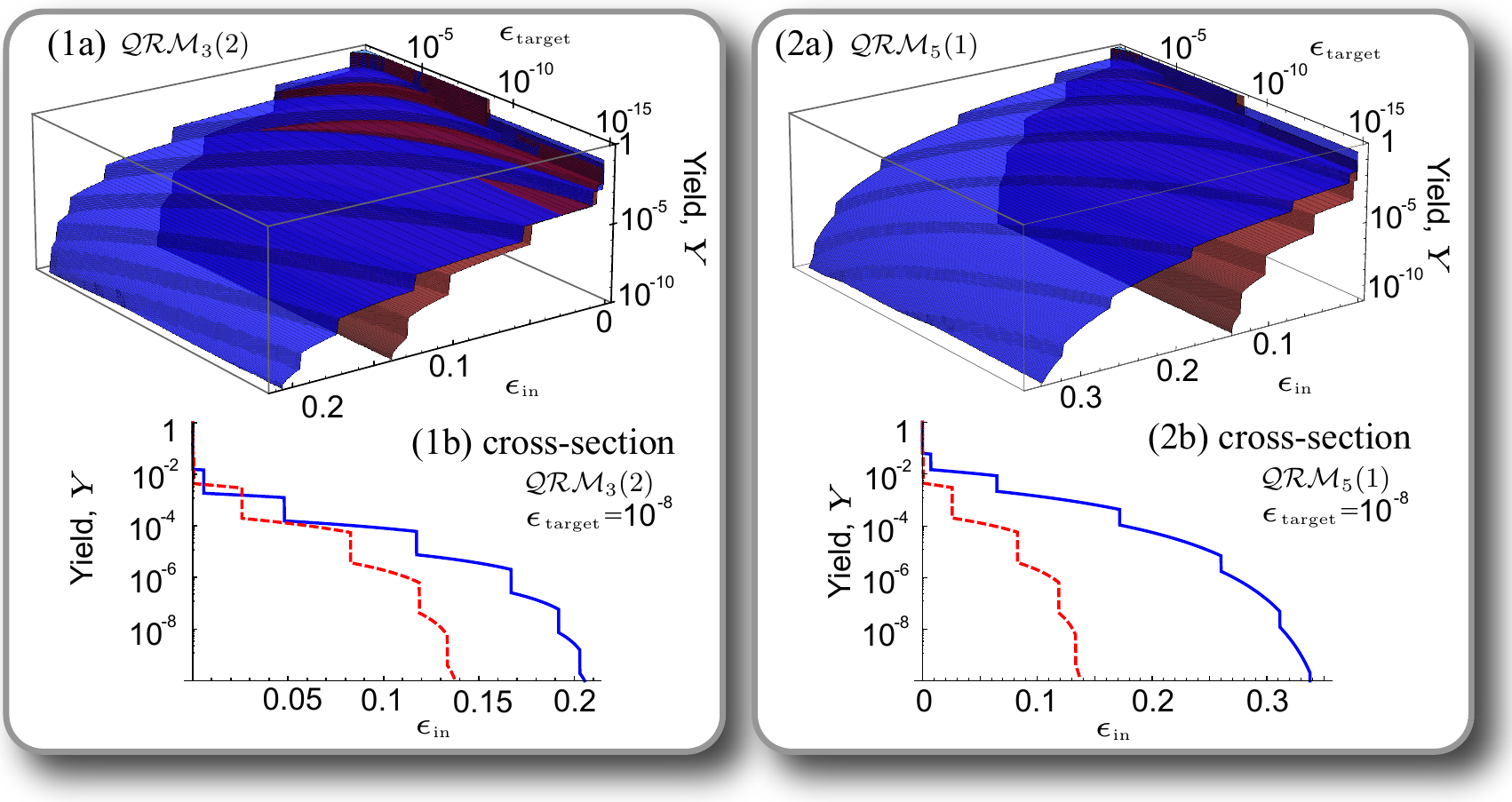}
\caption{The yield on a log-scale of of our protocols, $\mathcal{QRM}_{3}(2)$ and $\mathcal{QRM}_{5}(1)$, (blue) compared with the Bravyi-Kitaev $\mathcal{QRM}_{2}(4)$ (red) protocol.  Plots (1a) and (2a) are a function of initial error probabilities $\epsilon_{\mathrm{in}}$ and target error probabilities  $\epsilon_{\mathrm{target}}$.  For the qutrit and ququint states the noise is depolarizing.  Both (1a) and (2a) come with a cross-sections (1b) and (2b) respectively, with the target error probability held constant. The sudden changes in yield occurs because of discrete changes in the number of iterations required.}
\label{FidYield}
\end{figure*}

Unfortunately, 5-dimensional systems are quite complex.  Even after twirling into the $C_{M}$-plane, we cannot easily visually represent the whole distillability region as with did for the qutrit protocol.  For this reason we focus on the depolarized case with $f_{0}=1-\epsilon$ and $f_{j \neq 0} = \epsilon / 4$.  After a successful implementation of one round, a depolarized state is output with
\begin{eqnarray}
	\epsilon'  & = & \frac{\epsilon^2 (96 - 160 \epsilon + 75 \epsilon^2)} {64 - 256 \epsilon + 480 \epsilon^2 - 400 \epsilon^3 + 
 125 \epsilon^4}  ,  \nonumber \\
 & \sim & \frac{3 \epsilon^{2}}{2} + \frac{7\epsilon^{3}}{2}+ O[\epsilon^{4}] ,
\end{eqnarray}
and this occurs with probability
\begin{eqnarray}
	P & = & \frac{(1 - 2 \epsilon)^4 (64 - 256 \epsilon + 480 \epsilon^2 - 
   400 \epsilon^3 + 125 \epsilon^4)}{64 (-1 + \epsilon)^4} ,  \nonumber \\
   & \sim & 1 - 8 \epsilon + \frac{51 \epsilon^{2}}{2} + O[\epsilon^{3}] .
\end{eqnarray}
Based on these results we expect the protocol to have an excellent yield.  We numerically studied the yield  and again compared it against the qubit protocol $\mathcal{QRM}_{2}(4)$ or Bravyi and Kitaev, see Figs.~(\ref{FidYield}.2a,\ref{FidYield}.2b). The numerics confirm that across all parameter regimes  $\mathcal{QRM}_{5}(1)$ offers a significant resource savngs of potentially many orders of magnitude.  Magic state distillation is typically the most resource intensive aspect of fault tolerance schemes, and so high yield protocols are very desirable.

\section{State-injection and universal quantum computing}

Protocols for  qudit magic state distillation are our main focus, but what happens after preparation of a highly purified magic state?  Our ultimate goal is to simulate a non-Clifford group unitary via state-injection. For the $C_{M}$-magic states of direct interest we show the following.
\begin{theorem}
	Consider any $M \in \mathcal{M}_{d}^{m}$ and any noisy magic state $\sigma$ with $\epsilon =  1 - \bra{M_{0}} \sigma \ket{M_{0}}$.  There exists a trace preserving stabilizer operation, $\mathcal{G}$, that deterministically implements state-injection such that for all $\rho$
	\begin{equation}
		|| \mathcal{G}( \sigma \otimes \rho) - M \rho M^{\dagger} ||_{1} \leq 2 \epsilon.
	\end{equation}
	where $||..||_{1}$ is the trace norm $|| A ||_{1} = \tr ( \sqrt{A A^{\dagger}})$.
\end{theorem}
For perfect magic states, $\epsilon=0$, this entails that $ \mathcal{G}( \rho_{M} \otimes \rho) = M \rho M^{\dagger}$.  We first focus on the ideal case and later extend to noisy magic states.

For qubit systems, any magic state on the equator of the Bloch sphere may be exchanged for a unitary randomly selected from a pair of non-Clifford phase gates~\cite{BraKit05}.   In previous work, it was shown that a qutrit analog of the Bloch sphere equator~\cite{Hussain1} provides magic states that can be used for state-injection of non-Clifford phase gates.  Here we review and generalize these ideas.  
\begin{defin} We say a qudit quantum state $\ket{\Theta}$ is equatorial, or a phase state, if $\Theta \in \mathbb{R}^{d}$ and
\begin{equation}
	\ket{\Theta} =\frac{1}{\sqrt{d}} \sum_{j=1}^{d} e^{i \Theta_{j}} \ket{j} .
\end{equation}
\end{defin}
It follows immediately that a $\ket{M_{0}}$ state is a phase state with $\Theta_{j}=  2 \lambda_{j} \pi   / d^{m}  $.  The essential feature of such states is that they are unbiased with respect to the computational basis, such that a $Z$ measurement will generate completely random outcomes.  Taking an unknown state $\ket{\psi}$ and measuring $Z Z ^{\dagger}$ on the pair $\ket{ \Theta}\ket{\psi}$ will also give unbiased outcomes, and so no information is gained from $\ket{\psi}$.   Denoting a general state as $\ket{\psi} = \sum_{j} c_{j} \ket{j}$, the result of a projection, $\Pi_{k}$, onto a subspace stabilized by $\omega^{-k} ZZ^{\dagger}$ yields
\begin{equation}
	\Pi_{k} \ket{\psi}\ket{\Theta} \propto \sum_{j}  c_{j} e^{i\Theta_{j \oplus k}} \ket{j \oplus k }\ket{j} .
\end{equation}
We decode by performing a Clifford unitary such that $\ket{j \oplus k }\ket{j} \rightarrow \ket{k}\ket{j}$ and tracing out the first system.  As promised the result is a unitary transform, $\ket{\psi}\rightarrow U_{k}(\Theta)\ket{\psi}$, where
\begin{equation}
	U_{k}(\Theta)=\sum_{j} e^{i \Theta_{j \oplus k}} \kb{j}{j} ,
\end{equation}
which can also be expressed as 
\begin{equation}
	U_{k}(\Theta)= (X^{k})^{\dagger}U_{0}(\Theta) X^{k},
\end{equation}
where we note
\begin{equation}
	U_{0}(\Theta) \ket{+_{0}}= \ket{\Theta} .
\end{equation}
The transformation is unitary, but randomly selected from $d$ different possibilities.  

How do we simulate a deterministic unitary required for a computation?  Herein we consider unitary gates produced from a magic state, $\ket{M_{0}}$, such that
\begin{equation}
	U_{k} =  (X^{k})^{\dagger} M X^{k}.
\end{equation} 
Again we exploit the relationship between $M$ and the Clifford unitary $C_{M}=M X M^{\dagger}$.  Noting that $C_{M}^{k}=M X^{k} M^{\dagger}$ we express the unitary as
\begin{eqnarray}
	U_{k}& =&  (X^{k})^{\dagger} M X^{k} M^{\dagger} M , \\
	& = & (X^{k})^{\dagger} C_{M}^{k} M .
\end{eqnarray} 
Therefore, we can recover the desired $M$ unitary by applying the inverse of Clifford unitary $(X^{k})^{\dagger} C_{M}^{k}$.  

We have established a deterministic stabilizer operation, such that $\mathcal{G}( \kb{M_{0}}{M_{0}} \otimes \rho ) = M \rho M^{\dagger}$.  We now relax our assumptions and allow the resource to be imperfect, so that $1-\epsilon=\bra{M_{0}} \sigma \ket{M_{0}}$. By $C_{M}$-twirling we can ensure the state has the form $\sigma =(1-\epsilon)\kb{M_{0}}{M_{0}}+\epsilon \sigma'$ where $||\sigma'||_{1}=1$.  Applying our map $\mathcal{G}$ to the noisy state gives
\begin{equation}
	\mathcal{G}(\sigma \otimes \rho ) = (1-\epsilon)  M \rho M^{\dagger} + \epsilon \mathcal{G}(\sigma' \otimes \rho).
\end{equation}
Subtracting $ M \rho M^{\dagger}$ yields
\begin{equation}
	\mathcal{G}(\sigma \otimes \rho ) - M \rho M^{\dagger} =     \epsilon \mathcal{G}(\sigma' \otimes \rho) -\epsilon M \rho M^{\dagger} .
\end{equation}
Taking the trace norm and using the triangle inequality gives
\begin{eqnarray*}
	|| \mathcal{G}( \sigma \otimes \rho ) - M \rho M^{\dagger} ||_{1}&  \leq &     \epsilon || \mathcal{G} (\sigma' \otimes \rho) ||_{1} + \epsilon || M \rho M^{\dagger} ||_{1},  \\
 & = & 2 \epsilon .
\end{eqnarray*}
This is a rigorous treatment of the intuition that if the magic state is almost perfect then so too is the state-injection.

The addition of non-Clifford $M$ and $M^{\dagger}$ gates to our repertoire of unitaries generates a set dense in the special unitary group  (see Refs.~\cite{nrsbook,nrspaper} and App. \ref{universality}). Furthermore, for every gate in this set its inverse is also contained in the set. Thus the Solovay-Kitaev algorithm can be applied to ensure an efficient approximation of any unitary.  This argument also applies to the results of Ref~\cite{Hussain1}, where the qutrit Clifford group was supplemented by a non-Clifford unitary but universality was only conjectured there.

\section{discussion}

We have generalized the idea of magic state distillation using quantum Reed-Muller codes to all prime dimensions, enabling preparation of highly purified pure non-stabilizer states given a device capable of ideal stabilizer operations.  By state-injection these magic states enable us to simulate universal quantum computation.  While many aspects of the generalization were very analogous to the qubit case, there have also been some remarkable surprises.  In odd prime dimension the non-Clifford gates we gain are fundamentally different from the phase gates implemented by the Bravyi-Kitaev protocols.  In particular, we find that for primes $d \geq  5$ there exist quantum Reed-Muller codes of only $d-1$ qudits that possess these non-Clifford gates as transversal gates, whereas $2^{4}-1=15$ qubits are needed for a similar construction.  

To our knowledge the ququint code ($d=5$) using only 4 ququints is the smallest non-trivial stabilizer code with a transversal non-Clifford gate.  This translates into real practical gains, with the ququint protocol achieving better error probability thresholds (see Sec.~\ref{isoNoiseThresholds}) than any other known protocol with a polynomially scaling yield; $\epsilon^*_{\mathrm{dep}}=0.363$ for depolarizing noise.  Calculating the yield of the ququint protocol also shows that it is superior to all known qubit protocols, as demonstrated both by numerics and analytic scaling arguments.  For larger prime dimensions, $d > 5$, the thresholds and resource costs deteriorate with increasing dimension.  It is not presently clear whether this is an inevitable problem with higher dimensional systems or a peculiarity of our protocols.  We also investigate in detail the performance of an 8 qutrit ($d=3$) protocol, which whilst not as effective as the ququint protocol was still competitive against qubit protocols.  

It is natural to question whether $\epsilon$, as defined in equation (\ref{epsilon}), is a fair measure to compare noise thresholds in systems of different dimensions. It would be desirable to use a noise measure which is practically motivated based on a noise processes which could occur in the lab. In \cite{Howard11}, the depolarising noise rate $\delta$ is employed, where $\delta$ measures the degree of depolarising noise of state $\rho$ from pure state $\ket{\psi}$ state via $\rho=(1-\delta)\ket{\psi}\bra{\psi}+(p/d)\openone$. For a depolarising noise model $\delta$ is related to $\epsilon$ via $\epsilon=((d-1)/d)\delta$. Quantifying error via $\delta$ penalises higher dimensional states, yet even via this measure the thresholds for the 4-ququint code continue to significantly outperform their qubit counterparts. 

Nevertheless, one can argue that $\delta$ is also an unfair method of comparison given the larger number of noise processes which contribute to depolarising noise for higher dimensional systems. Ultimately, for the context of magic state distillation, the most relevant measure of comparison would be the yield at the fault-tolerance threshold. Unfortunately, at present, fault tolerant quantum computation with higher dimensional systems remains a little explored research area, and comparable thresholds to e.g. Knill \cite{Knill04} or Harrington et al \cite{Rauss07}'s schemes are unknown.  We know of only one study of qudit fault tolerance thresholds~\cite{Kanungo05}, and while evidence was presented that higher dimensional systems may provide better thresholds than their binary counterparts, the analysis in this paper is limited.  In particular therefore, our results motivate further study of an full fault-tolerance schemes based on ququint and qutrit components.  It is possible that the enhanced performance in dimensions 3 and 5 seen in our magic state distillation protocols translate into better thresholds and resource costs for full fault tolerance schemes based on qutrits and ququints. 

Another application of our results it to models of computation where the fault tolerant operations are a proper subgroup of the Clifford group.  For instance, the qubit topological cluster states~\cite{RHG01a,Rauss07} cannot directly prepare $Y$ eigenstates, but they can be distilled using magic state distillation.  In qudit generalizations of the topological cluster scheme, we anticpate that preparation of $XZ$ eigenstates will not be topologically protected.  While we have focused on distillation of non-stabilizer states, our protocols also enable distillation of $XZ$ stabilizer states.

Our understanding of the magic state model is still in its infancy despite many striking similarities with the more mature theory of entanglement.  However, as reviewed in the introduction there has been a flurry of recent results on qudit magic state model.  Numerous problems of a fundamental nature now present themselves as ripe for tackling.  Inspired by entanglement theory, we might ask if their exist qudit protocols for magic catalysis~\cite{Camp11,Catalysis99} or magic activation~\cite{Camp11,HoroBound99}.  Furthermore, while all known protocols offer yields of magic states with arbitrarily small error probabilities, the yield vanishes as the target error vanishes.  Contrast this with entanglement theory where the hashing protocol~\cite{BDSW01a,DevWinter} and quantum polar-coding techniques~\cite{PolarCode} offer a method of distilling entanglement at a non-zero yield even for vanishing target error.  Whether such a protocol could exist for magic state distillation is an intriguing and wide open question.

In the final stages of this research we became aware of recent work that proposes a novel protocol~\cite{Knill12} for qubit magic state distillation.  The protocol, which they call the 10-to-1 protocol, takes 10 noisy magic states each iterate and outputs 2 magic states.  This is the first protocol to output more than 1 magic state per iterate, and this has the benefit of increasing its yield.  Potentially similar techniques could also be used to design higher dimensional protocols.

\section{Acknowledgements}

We acknowledge the financial support of the EPSRC, the Leverhulme Trust, the BMBF (QuORep) and the EU (QESSENCE). DEB was supported in part by the Perimeter Institute, and would like to thank them for their hospitality. We would like to thank Martin Kliesch, Daniel Gottesman, Peter Shor, Joe Fitzsimons, Debbie Leung and Mark Howard for helpful discussions. In particular, we would like to thank Vadym Kliuchnikov for bringing~\cite{nrsbook} to our attention and for suggesting the argument given in Appendix \ref{universality}. 

\appendix

\section{The canonical M gate}
\label{QgateApp}

Here we verify the assertions of Thm.~\ref{thm_Mgate} and show that the canonical $M$ is a member of $\mathcal{M}_{d}^{m}$ for the asserted values of $d$Ê and $m$.    We begin by showing that
\begin{equation}
	C_{M}=M X M^{\dagger} \propto X P . 
\end{equation}
Left multiplying by $X^{\dagger}$ gives $ X^{\dagger} M X M^{\dagger} \propto P$.  The left hand side is then
\begin{equation}
	X^{\dagger} M X M^{\dagger} = \sum_{j} \exp(i 2 \pi  ( \lambda_{j \oplus 1} - \lambda_{j} ) / d^{m}) \kb{j}{j} .
\end{equation}
This equals $P$, upto a global phase, if for all $0 \leq j \leq q-1$,
\begin{equation}
\label{eqn_lambda_diff}
	\lambda_{j \oplus 1} - \lambda_{j} =d^{m-1}   \binom{j}{2} +  c  ,
\end{equation}
for some $c$. We first solve for the cases where $j \oplus 1 = j + 1$, that is $j \neq q-1$. For this set of equations, we may use standard arithmetic and recurrence equation methods, and the general solution is
\begin{equation}\label{appinter}
	\lambda_{j} = d^{m-1} \binom{j}{3}+j c + \lambda_{0} ,
\end{equation}
for all $j$, where $c$ and $\lambda_{0}$ are integers to be determined. These integer variables will be fixed by demanding that  Eq.~(\ref{eqn_lambda_diff}) with $j=d-1$ holds, and also that $\sum_{j} \lambda_{j}=0$. First, let us impose the former condition and substitute  Eq.~\eqref{appinter} into Eq.~\eqref{eqn_lambda_diff} for $j=d-1$, to yield  
\begin{equation}
\begin{split}
	\lambda_{0} - \lambda_{d-1} &=  \lambda_{0}-\bigl(d^{m-1} \binom{d-1}{3}+j(d-1) c + \lambda_{0}\bigr) , 
	\\ \nonumber &=d^{m-1}   \binom{j-1}{2} +  c .
\end{split}
\end{equation}
Solving this equation for $c$ yields
\begin{equation}
	c =  - d^{{m-2}}\binom{d}{3} .
\end{equation}
For $m\ge2$, inspection reveals that $c$ is integer-valued for all $d$. For $m=1$, $c$ is integer-valued for all prime $d\ge 5$. This follows from the fact that when $m=1$, $c=-(d-1)(d-2)/6$. We use the fact that $6=3\times 2$. Since $d\ge 5$ is a prime number not equal to three $d$ it is not divisible by three, thus either $(d-1)$ or $(d-2)$ must be divisible by three. Since  $d\ge 5$ is a prime number not equal to two then $(d-1)$ must be divisible by 2. Hence the product $(d-1)(d-2)$ is divisible by 6 for all primes $d\ge 5$, and $c$ is an integer for $m=1$ and $d\ge 5$.

It remains to fix $\lambda_{0}$ by imposing that $\sum_{j} \lambda_{j}=0$. Performing the summation and simplifying, we find that
\begin{equation}
\lambda_{0}= d^{m-2} \binom{d+1}{4} .
\end{equation}
Again, for $m\ge2$ this is (by inspection) integer-valued for all $d$. For $m=1$, this is integer-valued for all prime $d\ge 5$, and the proof for this latter case is similar to above. When $m=1$, $\lambda_{0}=(d+1)(d-1)(d-2)/24$. We observe that $24=3\times2\times4$. Since $d\ge 5$ is a prime number not equal to three $d$ it is not divisible by three, thus either $(d-1)$ or $(d-2)$ must be divisible by three. Since $d$ is an odd prime number both $(d+1)$ and $(d-1)$  must be divisible by two, and one of this pair must be divisible by 4. Hence $(d+1)(d-1)(d-2)$ must be divisible by 24 and consequently  $\lambda_{0}$ is an integer for $m=1$ and $d\ge 5$.

Thus, the gate $M$ as defined in theorem 1, satisfies all the requirements to be a member of $\mathcal{M}^{m}_{d}$. For $m=1$ and $d=3$, $\lambda_{j}$, is not integer-valued for all values of $j$ and so the above argument does not provide a member of $\mathcal{M}_{3}^{1}$.  Indeed, for $d=3$ it is easy to numerically search the sets of gates with integer $\lambda_{j}$ and verify that none are non-Clifford and so $\mathcal{M}_{3}^{1}$ is empty.

\section{Projection onto logical subspace}
\label{APP_projection}

Here we present the reasoning that leads to Eqs. (\ref{Eq_NoError},\ref{Eq_Error}), which can be divided up into 3 cases: a detected error, no error and an undetected error.

When $ \vec{v} \notin \mathcal{L}_{X}^{\perp}$, an error is present that is detected by the code and so the state vanishes, $\Pi \ket{+_{\vec{v}}}=0$.  To see this we recall that $X \ket{+_{k}}= \omega^{k} \ket{+_{k}}$ and so more generally $X[\vec{u}] \ket{+_{\vec{v}}}= \omega^{\langle \vec{v} , \vec{u} \rangle}  \ket{+_{\vec{v}}}$. Projecting onto the ``+1" eigenspace of all $X[\vec{u}] \in \mathcal{S}_{X}$ entails that the state will vanish unless $\langle \vec{v} , \vec{u} \rangle =0 $ for all $\vec{u} \in \mathcal{L}_{X}$.  This is simply the requirement that $\vec{v}$ is in the dual of $ \mathcal{L}_{X}$, which proves Eq.~(\ref{Eq_NoError}).

For the ``no error" instances, $\vec{v}  \in \mathcal{L}_{Z}$, the state does not vanish under projection.  Furthermore, since $\ket{+_\vec{v}}=Z[\vec{v}]\ket{+}^{\otimes n}$ and $\Pi Z[\vec{v}] = \Pi$ we have $ \Pi \ket{+_\vec{v}}=\Pi \ket{+}^{\otimes n}$ and so all such states must be projected onto the same logical state.  Finally, we observe that $\ket{+}^{\otimes n}$ is stabilized by $X_{L}=X^{\otimes n}$ and so $\Pi \ket{+}^{\otimes n} = \sqrt{c} \ket{+^{L}_{0}}$.  

All other possibilities correspond to undetected errors, resulting in a projection onto other logical states.  In such cases, $\vec{v} \in \mathcal{L}_{X}^{\perp}$ and so there must exist a $j \in \mathbb{F}_{d}$ such that $\vec{w}=\vec{v} \oplus j \vec{1} \in \mathcal{L}_{Z}$.  In terms of Pauli operators, we have $Z[\vec{w}]=Z[\vec{v}]Z[j \vec{1}]$ and so $Z[\vec{v}]=Z[\vec{w}]Z[(d-j)\vec{1}]$.  Since the logical operator is $Z_{L}=Z[(d-1)\vec{1}]$ it follows that $Z[\vec{v}]=Z[\vec{w}]Z_{L}^{j}$.  In terms of the quantum state, we have $\ket{+_{\vec{v}}}=Z[\vec{w}]Z_{L}^{j} \ket{+}^{\otimes n}$ and so after projection $\Pi \ket{+_{\vec{v}}} = \sqrt{c} Z_{L}^{j} \ket{+^{L}_{0}}=\sqrt{c} \ket{+^{L}_{j}}$.

\section{Weight enumerators}
\label{APPweightenum}

Here we find the weight enumerators for the shortened Reed-Muller codes $ \mathcal{L}_{X} = \mathcal{RM}^{*}_{d}(1,m) $ and $\mathcal{L}_{X}'= \mathrm{span}(\mathcal{L}_{X},\vec{1})$ as given in Eqs.~(\ref{weightLX}, \ref{weightLXone}).  Since $\mathcal{L}_{X}\subset \mathcal{L}_{X}'$ it is natural to start with $\mathcal{L}_{X}$ and then add the remaining terms.

First, $\mathcal{L}_{X}$ contains a zero vector $ (0,0,\dots,0) $ with zero Hamming weight. Second, all the remaining codewords, there are $d^{m}-1 $ such codewords, have $(d -1)$ zeros, i.e have Hamming weight $n-(d-1)=d^{m}-d $. Thus we have the weight enumerator
\begin{equation}
W_{\mathcal{L}_{X}}(x)=1+ (d^{m}-1)x^{(d^{m}-d)}.
\end{equation}
The enumerator for $\mathcal{L}_{X}'$ can be broken up into $d$ separate sums, since $\mathcal{L}_{X}' = \{ \mathcal{L}_{X } , \mathcal{L}_{X} \oplus \vec{1} ,... \mathcal{L}_{X} \oplus (d-1) \vec{1}  \}$, and so
\begin{eqnarray*} 
	W_{\mathcal{L}_{X}'}(x) & = & \sum_{j=0}^{d-1} 	W_{\mathcal{L}_{X} \oplus j \vec{1}}(x) , \\
	& = & W_{\mathcal{L}_{X}}(x) + \sum_{j=1}^{d-1} 	W_{\mathcal{L}_{X} \oplus j \vec{1}}(x)  .
\end{eqnarray*}
For the rest of this argument we focus on the $j \neq 0 $ terms. First, each $ j\mathbf{1} $ when added to the $ (0,0,\dots,0) $ vector will generate a codeword of full Hamming weight ( $n=d^{m}-1$).  Second,  each $ j \mathbf{1} $ when added to any other codeword of $ \mathcal{L}_{X} $ (other than the $ (0,0,\dots,0) $ vector) results in a codeword with $d^{m-1}$ zero's and so Hamming weight $n-d^{m-1}=d^{m}-1-d^{m-1} $.  For each $\mathcal{L}_{X} \oplus j \vec{1}$, there are $ d^{m}-1 $ such codewords and so
\begin{equation}
	W_{\mathcal{L}_{X} \oplus j \vec{1}}(x) = x^{(d^{m}-1 )} + (d^{m}-1)x^{(d^{m}-1-d^{m-1})}.
\end{equation}
For every $j \neq 0$ we get the same result and we have $d-1$ such sums, and so
\begin{eqnarray*} 
	W_{\mathcal{L}_{X}'}(x)	& = & W_{\mathcal{L}_{X}}(x) + (d-1) 	W_{\mathcal{L}_{X} \oplus  \vec{1}}(x) ,
\end{eqnarray*}
which expands out into the formula given in the main text with $x=\tilde{\mu}$.

\section{Adding any non-Clifford gate promotes the Clifford group to a universal set}
\label{universality}

We consider quantum circuits on $n$ qudits of odd prime dimension $d$. We will show here that the  combination of two theorems of Nebe, Rains and Sloane shows that the addition of any non-Clifford gate to the Clifford group generates a set of unitaries, that is dense in $\textrm{SU}(d^{n})$. In Ref.~\cite{nrspaper}, Theorem 7.3 implies that any finite group that contains the Clifford group must be generated by the Clifford group and a gate proportional to the identity. Thus the group $H$ generated by the Clifford group and a non-Clifford unitary (not proportional to the identity) cannot be finite and must be of infinite order.

In Corollary 6.8.2 of~\cite{nrsbook}, it is shown that any closed sub-group, $H$, satisfying $\mathcal{C}_{d}^{n} \subset H \subset \textrm{U}(d^{n})$,  must either be have finite order (ignoring global phase factors) or be $\textrm{SU}(d^{n})$. Combining this corollary with the previous theorem we conclude that the closure of the group generated by the Clifford group and any non-Clifford unitary (not proportional to the identity) is $\textrm{SU}(d^{n})$.

\end{document}